\documentclass[12pt,showpacs,twocolumn, preprintnumbers,aps,pre, reprint,superscriptaddress]{revtex4-1}

\usepackage{graphicx}
\usepackage{subfigure}
\usepackage{color}
\usepackage{float}
\usepackage{amsmath}
\usepackage{newtxtext,newtxmath}
\usepackage[linktocpage,colorlinks=true,linkcolor=blue,citecolor=blue,breaklinks=true]{hyperref}
\usepackage{verbatim}
\usepackage{breakcites}

\newcommand{\RN}[1]{%
  \textup{\uppercase\expandafter{\romannumeral#1}}%
}

\begin{document}

\preprint{}

\title{Sea-Ice Distribution and Mixed-Layer Depths in Fram Strait}

\author{Sahil Agarwal}
\email[]{sahil.agarwal@yale.edu}
\affiliation{Program in Applied Mathematics, Yale University, New Haven, USA}

\author{M. Grae Worster}

\affiliation{Institute of Theoretical Geophysics, Department of Applied Mathematics and Theoretical Physics, \\University of Cambridge, Wilberforce Road, Cambridge CB3 0WA, UK}

\date{\today}

\begin{abstract}

\begin{itemize}
\item Steady state location of ice edge determined locally.
\item Examined variation in ice edge between East Greenland Current and West Spitsbergen Current, with the sea-ice velocity being the controlling factor on its extent on annual time scales.
\end{itemize}

In an effort to understand the dynamics of the Arctic sea-ice edge, we present a simple model of heat and mass transfer in the Fram Strait that reveals some fundamental mechanisms controlling sea-ice extent in the marginal seas and the depth and properties of the Arctic mixed layer. We identify and study key mechanisms relating to the sea-ice wedge described by Untersteiner, a boundary-layer structure near the ice edge, demonstrating how ice thickness and extent depend on ice-export rates, atmospheric forcing and the properties of incoming warm and salty Atlantic water in the West Spitsbergen Current.  Our time-dependent results demonstrate a seasonal asymmetry between the rates of ice advance and retreat and explain the significant variations in the Southerly extent of sea ice across the Fram Strait, with a long ice tongue corresponding with the East Greenland Current. Our simple model indicates that thinning of the Arctic sea-ice cover will lead to warming and freshening of the North Atlantic, which would give a de-stabilizing feedback to the Arctic ice cover, leading to a slowdown of the Atlantic Meridional Overturning Circulation.

\end{abstract}


\maketitle



\section{Introduction}
\label{sec:intro}
There has been a dramatic decrease in the extent and thickness of Arctic sea ice over the past few decades, with multi-year ice giving way to mostly first-year ice in the present-day Arctic \citep{Stroeve:2007aa, Comiso:2008aa,OneWatt}. While aspects of this decline are simulated in climate models, it is useful to develop heuristic understanding of the mechanisms causing it with simpler, analytic models. The relatively thin Arctic sea ice is much more vulnerable to external forcing than the Antarctic polar cap, which sits on land and is a few kilometers thick. To leading order, the balance of heat fluxes across the ice to the atmosphere and the ocean controls its thickness, with the ice being much more sensitive to the variation in oceanic heat flux ($O$(1) $Wm^{-2}$) than the atmospheric heat flux ($O$(100) $Wm^{-2}$) \citep{Maykut:1971aa}.

The sea-ice cover in the Arctic, though perennial, has been thinning rapidly in the past few decades. Because this perennial ice cover requires the presence of Multi-Year Ice (MYI), i.e. ice that persists for more than one melt season, there have been discussions regarding when and if the Arctic can transition to a seasonal ice cover \citep{EW09, Tietsche:2011, SerrezeBarry:2011, MW:2011, MW:2012, Toppaladoddi:2017aa}. The debate { whether or not the Arctic has surpassed a \emph{tipping point}} has mainly arisen because of the decay of MYI cover that covered nearly two-thirds of the basin in the 1970s to less than one-third in the past decade \citep{OneWatt}, leading to a state where the Arctic is mostly covered by first-year ice. Since the first-year ice retreats rapidly, knowing the location of the ice edge is an important step in determining how external forcings influence the basin-wide sea-ice properties.

Bering Strait, the Canadian Archipelago and Fram Strait are some of the gateways through which the Arctic Ocean interacts with the rest of the world oceans \citep{Manley:1995aa, Woodgate:2005aa, Kwok:2004aa, Meredith:2001aa}. While all these gateways are important, Fram Strait is the largest, deepest and most significant \citep{Untersteiner:1988aa, Vinje:1986aa, Vinje:1998aa}. The amount of fresh water exchange and deep water formation in these regions influence global ocean circulations \citep{Aagaard:1968aa, Aagaard:1968ab, Aagaard:1989aa, Rudels:1991aa, Steur:2009aa}. {The sea-ice export  through the Fram Strait constitutes nearly 10\% of all the sea ice in the Arctic, with 60\% of this occuring in the winter and the rest during the summer \citep{Smedsrud:2017aa}.} \citet{Colony:1985}, used a simple stochastic model to study the statistics of residence time of sea ice in different sectors of the Arctic. They demonstrated that, while it takes nearly 5 years for sea ice in the Beaufort sector to either melt or exit through the Fram Strait, the time for termination in the marginal ice zones is only 1-2 years. They also showed that the ice at the North Pole has a 35\% probability that it will exit in 1 year through the Strait and 96\% that it will exit through the Strait sometime. The sea-ice edge in the Fram Strait forms the northern boundary of the Atlantic Meridional Overturning Circulation (AMOC) \citep{Mauritzen:1997aa, Jungclaus:2005aa, Koenigk:2006aa, Miles:2014aa}. The AMOC circulates the cold and fresh waters from the Arctic to the Antarctic and back, forming a circulation that regulates global climate.

\begin{figure*}[htbp]
    \centering
    \includegraphics[trim = 0 0 0 0, clip, width = \textwidth]{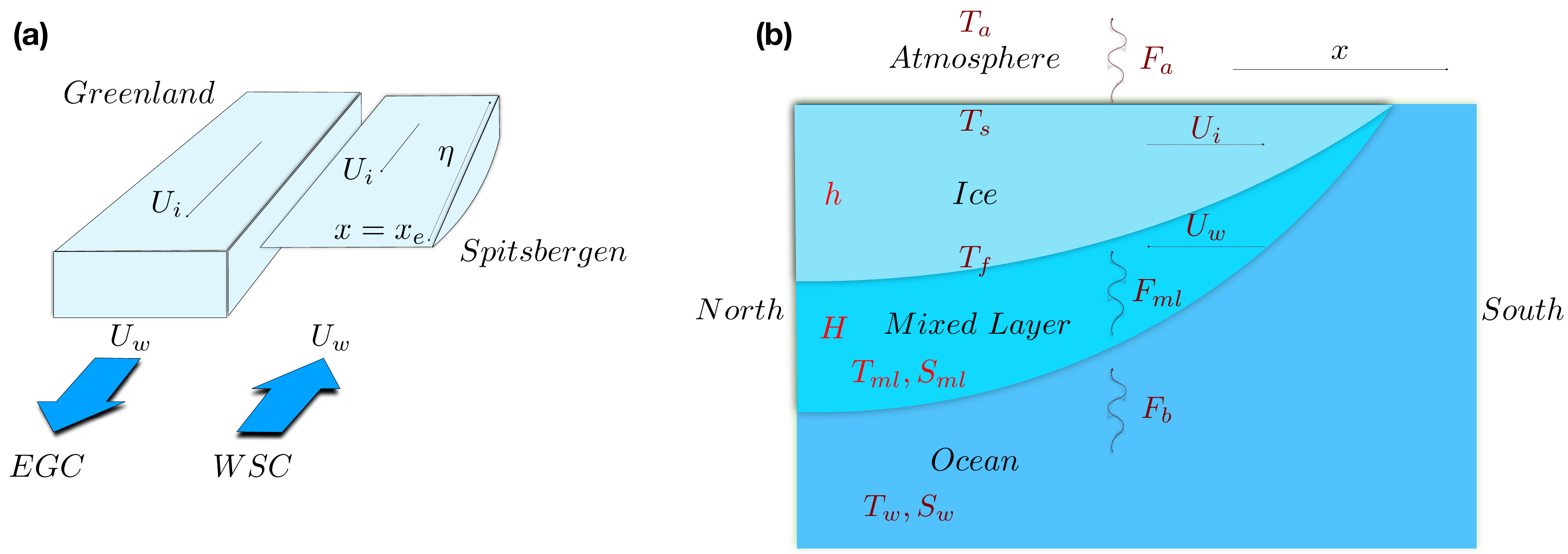}			  	
    \caption{(a) Schematic of an ice wedge, redrawn from Fig. 3 of \citep{Untersteiner:1988aa}. The warm and salty West Spitsbergen Current (WSC) flows northwards, which melts the southwards-flowing ice to create a wedge shape, whereas the relatively cold and fresh East Greenland Current (EGC) flows southwards with the ice and therefore does not generate such a wedge. (b) Schematic of our simple longitudinal model used to study the dynamics of the sea-ice edge in detail.}
    \label{fig:IceWedge}
\end{figure*}

\citet{Untersteiner:1988aa} studied the interaction of southwards-flowing ice with the northward-bound warm and salty waters of the West Spitsbergen Current \citep{Vowinckel:1962aa, Aagaard:1987aa}. The interaction of these two currents gives rise to an ice wedge, which can be imagined as follows: A rectangular slab of ice starts to melt as it comes in contact with warm water underneath. This melt water in turn is fresh and cold and starts to cool the incoming warm water (see Fig. \ref{fig:IceWedge}(a)), giving rise to an ice wedge, which we refer to as \emph{Untersteiner's Sea-ice Wedge}. In steady state, the heat energy required to melt the ice is comparable to the heat lost from the warm water current. \citet{Untersteiner:1988aa} computed several integral quantities over this wedge, such as the depth, temperature and salinity of the mixed layer in steady state, without calculating the length of the wedge itself or its detrended profile. { In reality the North Atlantic Warm Waters (NAWW) divide upon entering the Fram Strait into multiple branches, with a part circulating back in the EGC, another entering west of Spitsbergen, another branch dividing near the Barents sea and circulating cyclonically as a boundary current underneath the Arctic Ocean and returning under the EGC. In this study, as a simplification and as comparison to \citet{Untersteiner:1988aa}, we treat the WSC as the branch of the NAWW entering the Arctic west of Svalbard and ignore the rest of the branches. There are a multitude processes that happen in the Arctic such as formation of polynyas, river runoff from the Eurasian and American basins, cooling, freezing and melting of ice. Polynyas are one of most significant sources of saline rejection to mixed layer and deep waters in the Arctic, while river runoff brings fresh water to the Arctic. These water transformations in the complete Arctic basin lead to the boundary current of NAWW that enters the basin to exit near the Fram Strait as cool and fresh water, as the EGC. While the EGC can be thought of as an output of this complex non-linear large-scale dynamical system that is the whole Arctic, we treat the EGC and WSC as two individual currents and study their influence on the sea-ice formation and edge dynamics in the North Atlantic. }

Here, we study this phenomenon in more detail { by building a simple model for sea-ice export in the Fram Strait. This model allows us to analyze the individual physical mechanisms and their effect in much more detail than a complicated Global Circulation Model.} We look at the sea-ice-thickness distributions and mixed-layer properties along a longitude from a central basin in the Arctic to the sea-ice edge. Some of the key questions we ask are: What are the various factors responsible for formation of this ice wedge; Can we determine the ice edge based on external parameters, to know the location of the wedge; what is the typical length of the \emph{Untersteiner's Sea-ice Wedge} and what controls it; How do these ice-ocean interactions impact the dynamics of ice extent in the Arctic?

\section{Model}
\label{sec:CM}

Consider a slab of ice moving southwards from the north pole with velocity $U_i$ and an ocean current moving northwards with velocity $U_w$. The temperature and salinity of the oncoming ocean are considered to be constant at $T_w = 2^\circ{C}$ and $S_w = 36$\textperthousand, respectively. The relatively fresh, though cold melt water is less dense than the salty ocean water and forms a mixed layer beneath the melting ice (see Fig. \ref{fig:IceWedge}(b)).  The heat flux from the mixed layer to the underside of the ice $F_{ml} = \lambda_{ml} (T_{ml} - T_f)$, where $T_{ml}$ and $S_{ml}$ are the temperature and salinity of the mixed layer, $\lambda_{ml} = \rho c_p St |\bf{U_i} - \bf{U_w}|$ is the heat transfer coefficient \citep{McPhee:2008aa}, $c_p = 4186$ $J/kg/^\circ{C}$ is the specific heat capacity of water, and $St = 1.7 \times 10^{-4}$ is the Stanton number \citep{Omstedt:1992aa}. We assume that there is an abyssal heat flux from the ocean to the mixed layer $F_b$. In the presence of salt, the liquidus temperature $T_f$, which is decreased due to the presence of salt, can be approximated as $T_f = -\gamma S_{ml}$, where $\gamma \approx 0.055 ^\circ{C}/psu$. The net outgoing atmospheric heat flux at the surface of the ice is parameterized by $F_a = \lambda_a (T_s - T_a)$, where $T_s$ is the surface temperature at the ice surface, $T_a$ is the prescribed atmospheric temperature, and $\lambda_a$ is the heat-transfer coefficient, assumed to be a constant. Here, we take $T_a = -40 \cos \frac{\pi x}{2000}$ as a simple atmospheric forcing proxy to explore the role of latitudinally varying heat flux, where $x$ is distance in kilometers, measured southwards from the pole. The density of ice is assumed to be the same as the ocean ($\rho_{ice} = \rho_{ocean} = \rho = 1000kg/m^3$). For large \emph{Stefan number} ($= L/c_p \Delta T$), { where $\Delta T$ is the temperature difference between the top and bottom ice surfaces}, the heat-flux balance across the interfaces of ice of thickness $h$ and mixed layer of depth $H$ can be written as 
\begin{eqnarray}
&F_{b}& - F_a = -\rho L \frac{D^i}{D t} h + \rho c_p\frac{D^w}{D t} H(T_{ml} - T_w)
\label{eq:FbFa_NS}
\end{eqnarray}
for the total heat flux balance and
\begin{equation}
F_b - F_{ml} = \rho c_p \frac{D^w}{D t}H(T_{ml} - T_w)
\label{eq:FbFml_NS}
\end{equation}
for the balance across the mixed layer. Here, $L = 334 kJ/kg$ is the latent heat of fusion of ice, and $\frac{D^i}{D t}$ and $\frac{D^w}{D t}$ are material derivatives with respect to $U_i$ and $U_w$, respectively.

The fresh water from the ice melt is entrained by the mixed layer, thereby reducing the mixed-layer salinity. In other words, the salt content is conserved in the sea-ice -- mixed-layer system. Writing this salt flux balance for the whole system gives 
\begin{eqnarray}
\frac{D^i}{Dt}(S_i - S_w)h + \frac{D^w}{Dt} H(S_{ml} - S_w) = 0,
\label{eq:Salt}
\end{eqnarray}
where, $S_i$ is the salinity of ice. Assuming that the ice is fresh for simplicity, we take $S_i = 0$.

The mixed layer is assumed to deepen until it attains neutral buoyancy with respect to the underlying ocean to maintain a continuous density profile. This implies that the mixed layer has constant buoyancy $b$, which with an assumption of a linear equation of state gives
\begin{eqnarray}
\frac{D^w}{Dt}b &\equiv& \frac{D^w}{Dt}\left [ b_0 + (\alpha(T_{ml} - T_w) - \beta(S_{ml} - S_w) \right ] = 0. \nonumber \\
\Rightarrow
\alpha\frac{D^w}{Dt}T_{ml} &=& \beta\frac{D^w}{Dt}S_{ml}.
\label{eq:buoyancy}
\end{eqnarray}
where $\alpha$ and $\beta$ are coefficients of thermal expansion and solutal buoyancy, respectively.

The system of Eqns. \ref{eq:FbFa_NS}, \ref{eq:FbFml_NS}, \ref{eq:Salt} and \ref{eq:buoyancy} forms the complete system to be solved to determine sea-ice thickness $h$, mixed-layer depth $H$, mixed-layer temperature $T_{ml}$ and mixed-layer salinity $S_{ml}$. 

\subsection{Steady State}
\label{sec:SS}
In steady state, the coupled system of Eqns. \ref{eq:FbFa_NS} -- \ref{eq:buoyancy} can be written as
\begin{eqnarray}
\frac{\partial}{\partial x}H(T_w - T_{ml}) &=& \frac{\beta S_w(F_a - F_{ml}) }{\alpha L\rho U_w}, \nonumber \\
\frac{\partial}{\partial x} h &=& \frac{F_a - F_{ml}}{L \rho U_i}, \nonumber \\
\mathcal{F} = \frac{\alpha L (F_b - F_{ml})}{\beta c_p S_w (F_a - F_{ml})} &=& 1, \nonumber \\
\mathcal{B} = \frac{\alpha (T_w - T_{ml})}{\beta (S_w - S_{ml})} &=& 1.
\label{eq:solve2}
\end{eqnarray}
Here, $\mathcal{F}$ is the energy conservation parameter and $\mathcal{B}$ is the density ratio between the mixed layer and the ocean underneath. In the second equation above, the sea-ice melt occurs due to a mismatch between its bottom and top fluxes and hence our model is agnostic with regard to the location of the melt. Using the conditions that $\mathcal{F} = 1$ and $\mathcal{B} = 1$ (from Eqns. \ref{eq:FbFa_NS}, \ref{eq:FbFml_NS}, \ref{eq:Salt}, \ref{eq:buoyancy} and applying $L'H\hat{o}pital$'s rule, see Supporting Information (SI)) over the domain, we can obtain $T_{ml}$, substitute it back in the solution for the first equation above and obtain the mixed-layer depth $H$. Thus we find
\begin{eqnarray}
T_{ml} &=& \frac{K F_b - F_a}{\lambda_{ml}(K-1)} - \gamma S_{ml}, \qquad \textnormal{where } K = \frac{\alpha L}{\beta c_p S_w}.
\label{eq:Tml}
\end{eqnarray}
Hence,
\begin{eqnarray}
S_{ml} &=& S_w - \frac{\alpha}{\beta}(T_w - T_{ml}).
\label{eq:Sml}
\end{eqnarray}

This formulation reduces the coupled system of four equations to two equations for sea-ice thickness $h$ and heat content of the mixed layer $H(T_w - T_{ml})$, with the last two equations in (\ref{eq:solve2}) providing the conditions to determine the other properties.

\subsection{Time-dependent Periodic Forcing}
\label{sec:time}
We would also like to understand how these dynamics play a role in the sea-ice export in the Fram Strait, under an ideal periodic forcing of atmospheric temperature. A seasonal cycle is included by varying the atmospheric temperature $T_a$ by $10^\circ{C}$ over 365 days, i.e. $T_a(x,t) = (T_a(x) - 5) + 5 \cos \frac{2 \pi t}{365}$. 
The fully coupled time-dependent system of equations can be written as,
\begin{eqnarray}
\left(\frac{\partial}{\partial t} + {\bf{U_w}} \frac{\partial}{\partial x}\right)H(T_w - T_{ml}) &=& \frac{\beta S_w(F_a - F_{ml}) }{\alpha L\rho} \nonumber \\ \nonumber \\
\left(\frac{\partial}{\partial t} + {\bf{U_i}} \frac{\partial}{\partial x}\right) h &=& \frac{F_a - F_{ml}}{L \rho }\nonumber \\ \nonumber \\
\mathcal{F} = \frac{\alpha L (F_b - F_{ml})}{\beta c_p S_w (F_a - F_{ml})} &=& 1 \nonumber \\ \nonumber \\
\frac{\alpha (T_w - T_{ml})}{\beta (S_w - S_{ml})} &=& 1 \nonumber\\ \nonumber \\
sgn({\bf{U_w}}) &=& \left\{
     \begin{array}{@{}l@{\thinspace}l}
       -1  &: WSC\\
        1 &: EGC\\ 
     \end{array}
   \right \}
\label{eq:FullSystem}
\end{eqnarray}

The system of Eqs. \ref{eq:FullSystem} make up a simple model for mixed-layer properties and sea-ice thickness in the Fram Strait. We solve this hyperbolic system numerically. At first glance, one can see that this is a moving-boundary problem. Since the ice edge evolves in time, we map the domain onto a fixed rectangle using a new variable $\xi = x / x_e(t)$, where $0<x<x_e$ and $x_e$ is the ice edge at any time $t$. Substitution in to Eqs. \ref{eq:FullSystem} gives the equation governing the sea-ice thickness as
\begin{equation}
\frac{\partial}{\partial t}h - \frac{\dot{x_e}}{x_e} \xi \frac{\partial}{\partial \xi}h +  \frac{U_i}{x_e}\frac{\partial}{\partial \xi}h = \frac{F_a - F_{ml}}{L \rho}.
\label{eq:hscale}
\end{equation}
and the dynamics of the ice edge $x_e$ is given by
\begin{eqnarray}
\dot{x_e} &=& U_i, \qquad h_e > 0  \nonumber \\
\dot{x_e} &=& U_i - x_e \frac{\Delta F}{\frac{\partial h}{\partial \xi}}\big|_{\xi = 1}, \qquad h_e = 0.
\label{eq:xm}
\end{eqnarray}
where $\Delta F = \frac{F_a - F_{ml}}{L \rho}$ and $h_e$ is the ice thickness at the edge.

We solved the coupled system of Eqs. \ref{eq:FullSystem}, \ref{eq:hscale} and \ref{eq:xm} numerically using an upwind scheme. We prescribe a single ocean velocity profile across the Fram Strait based approximately on data from \citep{Aagaard:1968aa, Aagaard:1968ab, Foldvik:1988aa, Woodgate:1999aa, Cokelet:2008aa, Sutherland:2008aa, Steur:2009aa}  as Eq. \ref{eq:FS_Velocity} with distance measured west to east (with grid points $n$ every 21km), also depicted in Fig. \ref{fig:FS_VXM}(a), and keep the ice velocity fixed at $U_i = 0.1$ $ms^{-1}$. It has been shown that the waters in the Fram Strait do not interact significantly longitudinally in comparison to their latitudinal interaction, i.e. the measured east-west velocity is much smaller than the north-south component \citep{Boyd:1994aa, Woodgate:1999aa}. This allows us to use the model above to compute individual sea-ice profiles for each ($U_i,U_w$) velocity pair and stack them together. We limit the mixed-layer velocity in the East Greenland Current $U_w \le 0.075$ $ms^{-1}$ by assuming that the flow of the mixed layer is controlled by the flow of sea ice above it.

\begin{eqnarray}
{\hspace{-15mm}
U_w = \left\{
     \begin{array}{@{}l@{}ll}
     	\text{ EGC}&&\\
       0.075 \text{ ms}^{-1} &: 0km - 105km, &1 \le n \le 6,\\
	
	\text{EGC to WSC}&&\\
       0.075\cos \left(\frac{\pi (n-6)}{20} \right) \text{ ms}^{-1} &: 105km - 525km, &7 \le n \le 25,\\ 
	
	\text{ WSC}&&\\
       -0.257 \text{ ms}^{-1} &: 525km - 630km, &26 \le n \le 31,\\
     \end{array}
   \right.
   \label{eq:FS_Velocity}
   }
\end{eqnarray}

\section{Discussion and Results}
\label{sec:results}
\subsection{Steady State Results}
{Steady-state dynamics help in explaining how different components of the model interact with each other under a constant forcing. In addition to the longitudinal profiles for sea-ice thickness, and mixed layer properties, our simple model also allows us to compute the location of the sea-ice edge, the basin properties, and a wedge length scale analogous to \emph{Untersteiner's sea-ice wedge} analytically. }

\subsubsection{Analytic Solution for the Location of Ice Edge}
This simple model allows us to compute the location of the ice edge analytically, based on external parameters, which can be more easily observed.

From  Eq. \ref{eq:solve2}, $\mathcal{F} = 1$ at all places wherever there is ice, and hence $\mathcal{F}_{edge} = 1$ ($\mathcal{F}$ at edge). Therefore using $\mathcal{F}_{edge} = 1$ we can obtain the location of the steady-state ice edge $x_e$ from
\begin{eqnarray}
F_{a_{edge}}(x_e) &=& \frac{\alpha L}{\beta c_p S_w} (F_b - F_{ml_{edge}}) + F_{ml_{edge}},
\label{eq:Edge}
\end{eqnarray}
where $ F_{a_{edge}}$ and $F_{ml_{edge}}$ are the atmospheric and mixed-layer heat fluxes at the ice edge, respectively. Because the boundary conditions for temperature and salinity at the ice edge are known externally ($F_{a_{edge}}(x) = -\lambda_a (\gamma S_w + T_a(x))$,  $F_{ml_{edge}} = \lambda_{ml}(T_w + \gamma S_w)$), we can substitute those back in Eq. \ref{eq:Edge} and solve for $T_{a_{edge}}$. Since $T_a(x)$ is prescribed, the location of the ice edge $x_e$ is the root of the equation 
\begin{equation}
T_a(x_e) = T_{a_{edge}}.
\label{eq:TaEdge}
\end{equation}

The sea-ice edge in Fig. \ref{fig:TmlH_h} and Figs. S1-S2 in the SI were computed using this analysis and then the system was integrated northwards to get the sea-ice profile and mixed-layer properties for the complete domain.

\subsubsection{Basin Properties}

The basin properties in this idealized model can be computed analytically as follows. The Arctic basin in steady state can be assumed to have $F_b = F_{ml}$ and $F_b = F_a$, i.e. the abyssal heat flux from the ocean is fully transmitted through the mixed layer and ice to the atmosphere, which gives
\begin{eqnarray}
F_b = \lambda_{ml} (T_{ml} + \gamma S_{ml}), \textnormal{ and}
\label{eq:Basin2}
\end{eqnarray}
\begin{eqnarray}
F_b &=& \lambda_a (T_s - T_a) \nonumber \\
\Rightarrow T_s &=& \frac{F_b}{\lambda_a} + T_a.
\label{eq:Ts2}
\end{eqnarray}

Substituting $S_{ml}$ from Eq. \ref{eq:Sml} in Eq. \ref{eq:Basin2} gives us an equation for $T_{ml_{Basin}}$
\begin{equation}
T_{ml_{Basin}} = \frac{F_b - \gamma\lambda_{ml} \left ( S_w - \frac{\alpha}{\beta}T_w \right )}{\lambda_{ml} \left ( 1 + \gamma\frac{\alpha}{\beta} \right )}.
\label{eq:Basin3}
\end{equation}
Substitution of $T_{ml_{Basin}}$ from Eq. \ref{eq:Basin3} in Eq. \ref{eq:Sml} gives us $S_{ml_{Basin}}$. For large \emph{Stefan number} ($= L/c_p \Delta T$), the temperature profile in the ice can be assumed to be linear, which gives
\begin{equation}
T_s = \frac{-k\gamma S_{ml} + \lambda_a h T_a}{k + \lambda_a h}
\label{eq:Ts1}
\end{equation}
where $k = 2.2 W m^{-1} K^{-1}$ is the heat conductivity of ice. The thickness of sea ice in the basin is obtained by equating Eqs. \ref{eq:Ts1} and \ref{eq:Ts2}, which gives
\begin{equation}
h_{Basin} = \frac{-k \left ( \gamma S_{ml} + T_a(0) + \frac{F_b}{\lambda_a}\right )}{F_b}
\label{eq:hBasin}
\end{equation}

The assumption of $F_b = F_{ml}$ and $F_b = F_a$ in the Arctic basin is similar to the assumption by \citet{Colony:1985} of no ice formation in the Central Arctic. In other words, since ice is always present in the basin, it requires that the rate of formation of ice must always be greater than the ice velocity $U_i$. This strict assumption holds in Figs. S1 and S2, where we show both $F_a = F_{ml} = F_b = 20 Wm^{-2}$. $T_{ml_{Basin}}$, $S_{ml_{Basin}}$ and $h_{Basin}$ are marked in Figs. S1 and S2, which show very good agreement with this analysis.

\subsubsection{Untersteiner's Sea-ice Wedge}

\begin{figure}[htbp!]
	\includegraphics[trim = 0 0 0 0, clip, width = 0.5\textwidth]{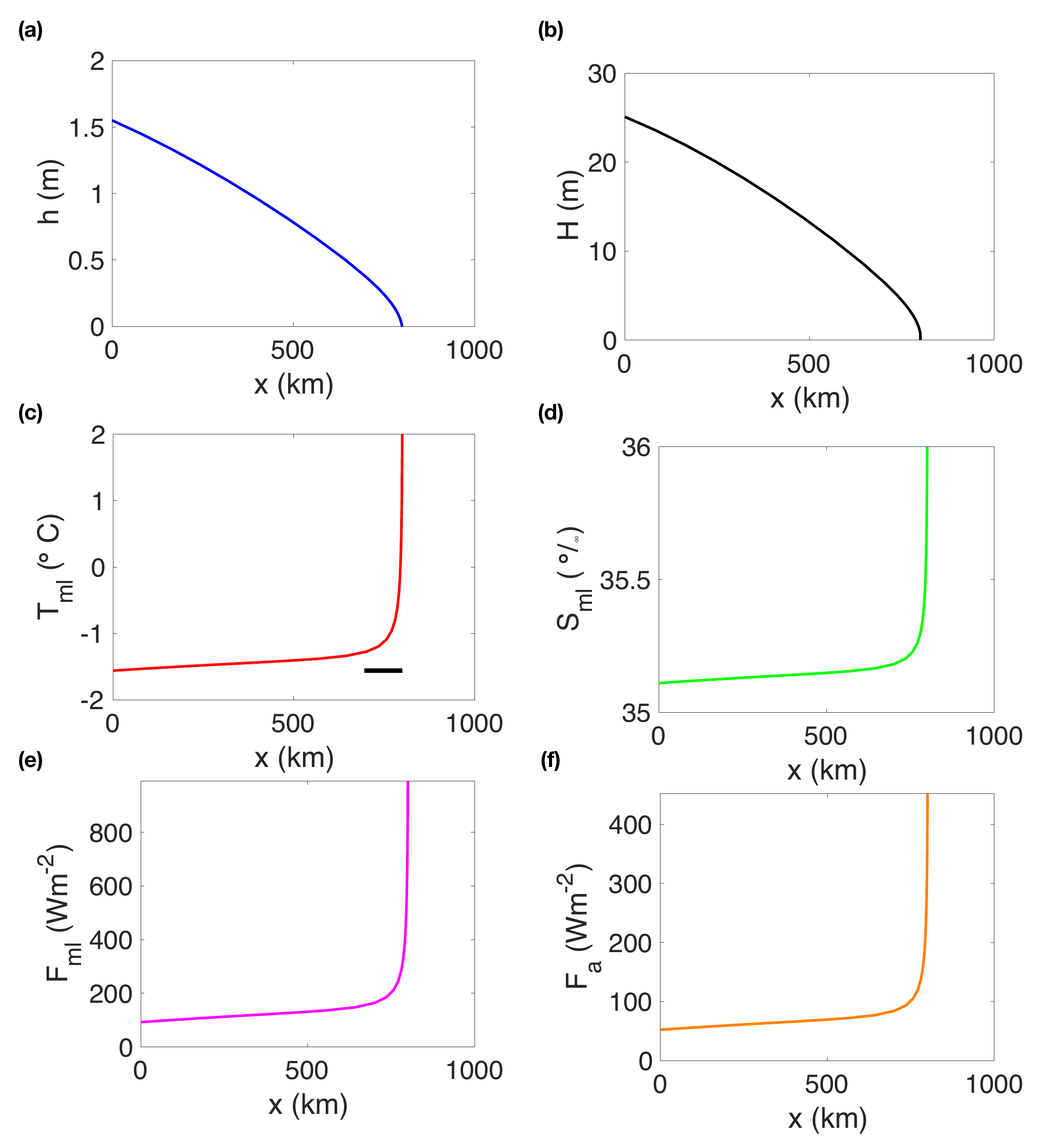}	
    \caption{With ice velocity $10^{-1}m/s$ and ocean velocity $2.5\times10^{-1}m/s$.(a)Ice thickness, $h$;(b)mixed-layer (ML) depth, $H$;(c)ML Temp., $T_{ml}$, Thick black line is the \emph{Untersteiner's Sea-ice Wedge}, $\eta = 105.5km$;(d)ML Salinity., $S_{ml}$; (e)mixed-layer heat flux $F_{ml}$; (f) Atmospheric heat flux $F_a$}
    \label{fig:TmlH_h}
\end{figure}

Imagine a rectangular block of sea ice moving southwards. As it starts to interact with a northward-bound warm and salty ocean current, it starts to melt \citep{Untersteiner:1988aa}. In steady state, it takes a wedge shape, with the heat required to melt the ice being equal to heat released from cooling of the incoming warm waters (see Fig. \ref{fig:IceWedge}(a)). A length scale $\eta$ can be ascribed to this wedge related to this transfer of heat. The maximum amount of heat being advected with respect to ice is

\begin{equation}
F_{Advect} = \rho c_p |U_w | H_o (T_{ml} - T_f)
\label{eq:F1}
\end{equation}
where $H_o$ is the mixed layer depth in the basin, calculated numerically. The amount of heat transferred from the mixed layer to ice that results in the melting of ice is
\begin{equation}
F_{Transfer} = \rho c_p St |U_i - U_w| \eta (T_{ml} - T_f).
\label{eq:F2}
\end{equation}
Here, $\eta$ is the length of the wedge over which $F_{Advect} = F_{Transfer}$. Equating Eqs. \ref{eq:F1} and \ref{eq:F2}, we get
\begin{equation}
\eta = \frac{|U_w| H_o}{|U_i - U_w| St}.
\label{eq:wedge}
\end{equation}
We show this wedge length scale in the solutions in Fig. \ref{fig:TmlH_h}.

We plot the solutions to Eqns. \ref{eq:solve2} in Fig. \ref{fig:TmlH_h} for $U_i = 10^{-1}m/s, U_w = 2.5\times10^{-1}m/s$. Fig. \ref{fig:TmlH_h} shows that the \emph{Untersteiner Sea-ice Wedge} is more apparent in the mixed-layer properties than in the sea-ice profile itself whose shape is determined dominantly by the latitudinal variation in atmospheric forcing. This wedge forms a boundary layer between the incoming ocean current and the mixed layer generated by the melting of ice. The wedge length scale in \citep{Untersteiner:1988aa} was approximated as 150 km, by arguing that melting ice is still apparent at 100 km from the ice edge, whereas at 200 km there is no sign of further melting. Using similar velocity profiles and external parameters, we calculate the wedge to have a length of approximately 105 km (see Fig. \ref{fig:TmlH_h}(c)). Untersteiner assumed $T_w = 2 ^\circ{C}$, $S_w = 35$\textperthousand, $U_i/U_w = 1/2$, and the sea-ice thickness at the end of the wedge to be 2.5m which gave a mixed-layer thickness at the end of the wedge of 24m, with mixed-layer temperature $T_{ml}$ of $-1.74^\circ{C}$. Given approximately the same parameters, we calculate $T_{ml}$ in this model as $-1.56^\circ{C}$. The average heat flux from the mixed layer to ice $F_{ml}$ was approximated as $300$ $Wm^{-2}$ in \citep{Untersteiner:1988aa}, while we calculate this quantity as $\sim 400$ $Wm^{-2}$. { To compare our model with observations, we used the mean atmospheric temperature profile over the Arctic from the \emph{Berkeley Earth Temperature} database \citep{Rohde:2013aa} over the period 1951-1980. This gives us $h = 2.43m$, $T_{ml} = -1.73^\circ{C}$ and $H = 37m$ in the Central Arctic, in agreement with the observed quantities during that period. }

We vary the ocean current and sea-ice velocities to study the effect of $U_i$ and $U_w$ on these profiles. As the ice velocity is decreased, keeping the ocean current velocity fixed, the ice wedge becomes sharper (see Fig. S1 in SI, $U_i = 10^{-3}m/s, U_w = 2.5\times10^{-1}m/s$). As both ice and ocean currents are slowed down, the ice wedge becomes extremely narrow, behaving more like the rectangular ice slab that flows south (see Fig. S2 in SI,  $U_i = 10^{-5}m/s, U_w = 2.5\times10^{-2}m/s$). Fig. S4 in SI compares four velocity pairs here with the situation if there were only atmospheric forcing. This shows the sensitivity of sea-ice thickness to the heat flux from the mixed layer as well as the formation of the \emph{Untersteiner Sea-ice Wedge.}

\begin{figure}[htbp]
    \centering
    \includegraphics[trim = 0 0 0 0, clip, width = 0.5\textwidth]{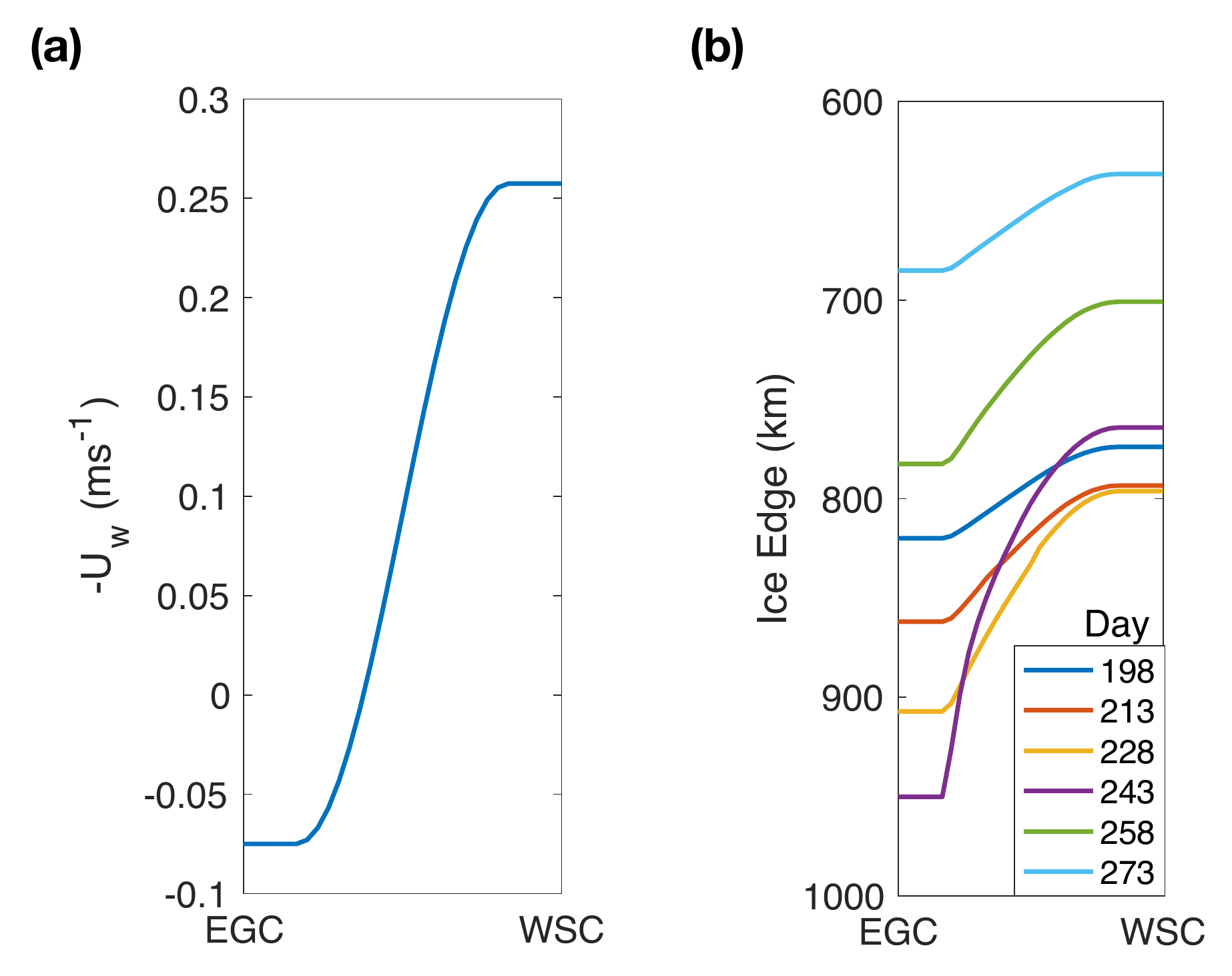}
    \caption{(a)Velocity profile of the mixed layer in the Fram Strait from Eq. \ref{eq:FS_Velocity}. (b) Location of the ice edge in the Fram Strait as a function of time. The ice edge extends for more than 200km in the East Greenland Current compared to the West Spitsbergen Current.}
    \label{fig:FS_VXM}
\end{figure}

\subsection{Time-dependent Forcing}

To understand how a changing climate affects the sea-ice profile and its associated properties, one needs to look at the fully time-dependent dynamics because, depending on the rate of change of climate forcing, the system may never reach a steady state and will always be in a state of evolution. In Fig. \ref{fig:FS_VXM}(b) we plot the location of the ice edge in the Fram Strait as a function of time, where day number denotes the time from the start of the simulation. Fig. \ref{fig:FS_VXM}(b) shows the asymmetry in the slow growth rate of ice as it moves southwards vs the fast melt rate as it retreats northwards. It takes more than 45 days to grow approximately 150km while it takes less than 30 days to retreat back more than 250km. The difference in extent between the East Greenland Current vs. the West Spitsbergen Current is more than 200km, determined only by the variation in the ocean-current velocity. Another point to note is the difference in time when the respective longitudes reach the maximum ice-extent. The ice-extent in EGC keeps increasing even when the ice-edge in the WSC has started to retreat. This lag produces a nonlinear response to an almost linear change in the relative velocity. \citet{Agarwal:2017aa} showed that sea-ice velocity may be the dominant physical process in explaining the variability in the ice-extent dynamics in the Arctic on timescales up to a few years. This independent modeling study shows that only a small change in the relative velocity between ice and ocean currents can cause a very large variation in the sea-ice extent. We show the time series of the sea-ice edge for two velocity profiles in Fig. S5. These time series also show the transient behavior that the system exhibits to come to a steady seasonal cycle. The entire simulation showing the daily sea-ice edge for two seasonal cycles is shown in Movie S1. { To compare this model with the observations, similar to the steady state case, we used the \emph{Berkeley Earth Temperature} database \citep{Rohde:2013aa}. Varying the ocean velocity from 0.095$ms^{-1}$ in the EGC to 0.25$ms^{-1}$ in the WSC produces a sea-ice tongue in excess of 600km, showing the robustness of this model in studying the individual mechanisms of sea-ice export.}

\subsection{Implications for the Atlantic Meridional Overturning Circulation}
\begin{figure*}[htbp!]
	\includegraphics[trim = 0 0 0 0, clip, width = \textwidth]{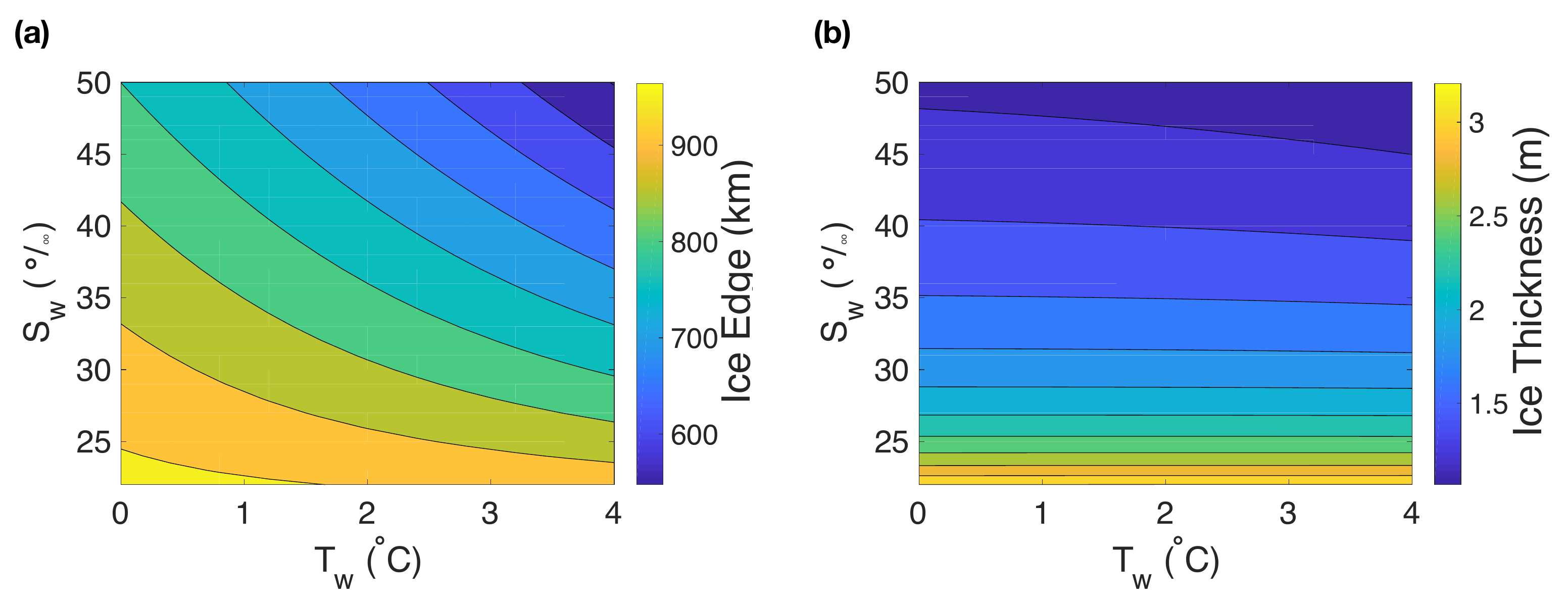}	  	
    \caption{Steady state location of the sea-ice edge and basal sea-ice thickness as a function of $T_w$ and $S_w$ of the incoming Atlantic waters through the WSC, with $F_b = 20 Wm^{-2}$.}
    \label{fig:SS_V}
\end{figure*}

Sea ice is a major component of the Atlantic Meridional Overturning Circulation (AMOC), which regulates global climate by transporting heat and salt across the global oceans. Southwards flowing sea ice  and the relatively fresh mixed layer waters from the EGC provide a net freshwater flux to the AMOC, thus modulating its strength on timescales of decades and more \citep{Aagaard:1989aa, Timmermann:2003aa,  Sevellec:2017aa}. The location of the Arctic sea-ice edge can have huge influence on the strength of the AMOC \citep{Sevellec:2017aa}. Figs. \ref{fig:SS_V}(a,b) show the steady state location of the ice edge and basal sea-ice thickness as the temperature and salinity of the incoming Atlantic waters through the WSC are varied. These calculations show that the basal ice thickness does not vary strongly with the temperature of WSC, but only with its salinity. On the other hand, the location of the sea-ice edge has a strong dependence on both the temperature and salinity of the WSC. These are in agreement with the recent warming of the incoming Atlantic Waters through the WSC \citep{Walczowski:2007aa,Schauer:2008aa}  accompanied by the migration of the sea-ice edge northwards in the Greenland Sea leading to a decrease in the sea-ice extent. 
Movie S2 shows the simulation of sea-ice flow across the Fram Strait, with the thickness variation shown in colors. To calculate the annual sea-ice flux across the Fram Strait, we use the thickness profile across the north-most ice edge, which is a consequence of the WSC entering the Arctic near Spitsbergen. The volume flux $F_v$ is calculated as $F_v = \int_{n=1}^{n=31} U_n h_n dx$, where $U_n = 0.1ms^{-1}$ is the ice velocity, $h_n$ is the ice thickness at grid point $n$ and $dx = 21km$ is the spacing between each grid point. We compute this integral using the trapezoidal rule, which gives a value for $F_v = 1234km^3/year$ or about $0.04 Sv$, comparable to the noise amplitude used in simulations for determining the stability of the thermohaline circulations \citep{Timmermann:2003aa}. Figure S6 shows the daily sea-ice flux across the Fram Strait. \citet{Aagaard:1989aa} give a value of $2790km^3/year$ as an estimate of the sea-ice flux to the Greenland-Iceland-Norwegian Seas. While our value for the idealized forcings is significantly lower than their value, there are a few important reasons for this difference: (a) we use a constant sea-ice velocity profile, with $U_i = 0.1m/s$, while the realistic velocities may be larger in the EGC; (b) we use a very simplified ocean velocity profile across the Fram Strait (see Fig. \ref{fig:FS_VXM}(a)); (c) these computations are only looking at the first-year ice, which is considerably thinner than multiyear ice; {(d) the atmospheric temperature profile we assume is much simpler than observed. }
Retreat of the Arctic sea ice has led to increased warming and freshening of the open ocean in the North Atlantic with the warming having a greater impact than the freshening, leading to a destabilizing feedback to the Arctic sea ice. This retreat in turn gives rise to buoyancy anomalies in the North Atlantic causing a slowdown of the AMOC  \citep{Sevellec:2017aa} and hence affecting the global climate dynamics.

\section{Conclusions}
\label{sec:conc}
Fram Strait is the largest gateway in the Arctic through which the polar ocean interacts with the world oceans \citep{Untersteiner:1988aa}, and the sea-ice edge in the Fram Strait forms the northern boundary for the formation of Deep Water in the North Atlantic that drives the Atlantic MOC. Therefore, the location of the sea-ice edge is a factor influencing the strength and dynamics of this circulation, which in turn influences the global climate through ocean circulation, with time scales ranging from decades to centuries.

\citet{Thorndike:1982} have shown that more than $70\%$ of the variability in the sea-ice velocity fields can be explained by the geostrophic winds. \citet{Agarwal:2017aa} have shown further that this variability in the sea-ice velocity fields explains the dynamics of the observed sea-ice extent on time scales ranging from a few days up to a couple of years. Using a simple model, we have shown how the interaction between sea-ice currents flowing southwards and  ocean currents flowing southwards/northwards (East Greenland Current/WestSpitsbergen Current) influences the sea-ice edge and therefore sea-ice extent in the Fram Strait. { While the idealized forcings produce a sea-ice tongue of approximately 200km, the climatological temperature profile obtained from observations leads to a sea-ice tongue on the order of 600km, in agreement with the observations.}

{ The Atlantic waters entering the Arctic have become warmer as compared to the past few decades \citep{Walczowski:2007aa,Schauer:2008aa}, leading to an increased melt of sea ice and the thinner ice, being more susceptible to wind forcing, has become faster when flowing out of the Arctic. Our simple model has allowed us to vary parameters such as the ice and ocean current velocities, Atlantic temperature and salinity as well as atmospheric temperature to study their effect on the dynamics of sea-ice edge and accompanying mixed layer properties in the Fram Strait.} We show how the interaction between the oncoming sea-ice and ocean currents form a boundary-layer structure with a length scale associated with the \emph{Untersteiner Sea-ice Wedge}, which is more apparent in the mixed-layer properties than the sea ice itself. Our simple model also shows how to calculate values of key variables such as the sea-ice edge only from the external forcing, without integrating the complete model. This allows us to study the effect of climate change on the Arctic sea-ice extent directly. The dynamics of sea-ice export in the Fram Strait is a highly nonlinear coupled complex dynamical system, which produces a nonlinear response given a nearly linear velocity change, which is exhibited by our time-dependent solutions that show asymmetrical growth and melt times.

Our simple model is able to capture an ice tongue along Eastern Greenland that extends more than 200km further southwards than the Spitsbergen side of the system. It should be borne in mind that our numerical solutions only involve the export of first-year ice from the Arctic basin, which is then melted as it flows southwards, which produces a \emph{shorter} ice tongue than observed.  A key physical ocean phenomenon prevalent near the sea-ice edge are the eddies. While our simple model does not include these highly complicated flows explicitly, their overall physics is subsumed by the imposition of sea-ice / ocean velocity or the strength of the ocean current which governs the heat flux at the boundaries. There are many other physical mechanisms that have not been included in our simple model above, such as the Coriolis effect on the western boundary current and fresh-water runoff from Greenland, to name a few. However, this simple model illustrates important effects of heat and mass transfer between ice and ocean that contribute dominantly to the formation of the Arctic mixed layer and influence the principal physics of the marginal ice zone and the extent of exported sea ice. It also provides a framework to study the interaction of ocean and sea-ice currents in greater detail.

\acknowledgments
MGW is grateful for support from the Yale Institute for Biospheric Studies. This work started at the 2016 Geophysical Fluid Dynamics Summer Program, Woods Hole Oceanographic Institution, which is supported by the National Science Foundation and the Office of Naval Research. We thank all the Fellows and staff for their useful discussions. We also thank J.S. Wettlaufer for his comments. SA acknowledges support from the David Crighton Fellowship at University of Cambridge where this work was continued and Yale University for financial support.


\begin{thebibliography}{44}%
\makeatletter
\providecommand \@ifxundefined [1]{%
 \@ifx{#1\undefined}
}%
\providecommand \@ifnum [1]{%
 \ifnum #1\expandafter \@firstoftwo
 \else \expandafter \@secondoftwo
 \fi
}%
\providecommand \@ifx [1]{%
 \ifx #1\expandafter \@firstoftwo
 \else \expandafter \@secondoftwo
 \fi
}%
\providecommand \natexlab [1]{#1}%
\providecommand \enquote  [1]{``#1''}%
\providecommand \bibnamefont  [1]{#1}%
\providecommand \bibfnamefont [1]{#1}%
\providecommand \citenamefont [1]{#1}%
\providecommand \href@noop [0]{\@secondoftwo}%
\providecommand \href [0]{\begingroup \@sanitize@url \@href}%
\providecommand \@href[1]{\@@startlink{#1}\@@href}%
\providecommand \@@href[1]{\endgroup#1\@@endlink}%
\providecommand \@sanitize@url [0]{\catcode `\\12\catcode `\$12\catcode
  `\&12\catcode `\#12\catcode `\^12\catcode `\_12\catcode `\%12\relax}%
\providecommand \@@startlink[1]{}%
\providecommand \@@endlink[0]{}%
\providecommand \url  [0]{\begingroup\@sanitize@url \@url }%
\providecommand \@url [1]{\endgroup\@href {#1}{\urlprefix }}%
\providecommand \urlprefix  [0]{URL }%
\providecommand \Eprint [0]{\href }%
\providecommand \doibase [0]{http://dx.doi.org/}%
\providecommand \selectlanguage [0]{\@gobble}%
\providecommand \bibinfo  [0]{\@secondoftwo}%
\providecommand \bibfield  [0]{\@secondoftwo}%
\providecommand \translation [1]{[#1]}%
\providecommand \BibitemOpen [0]{}%
\providecommand \bibitemStop [0]{}%
\providecommand \bibitemNoStop [0]{.\EOS\space}%
\providecommand \EOS [0]{\spacefactor3000\relax}%
\providecommand \BibitemShut  [1]{\csname bibitem#1\endcsname}%
\let\auto@bib@innerbib\@empty
\bibitem [{\citenamefont {Stroeve}\ \emph {et~al.}(2007)\citenamefont
  {Stroeve}, \citenamefont {Holland}, \citenamefont {Meier}, \citenamefont
  {Scambos},\ and\ \citenamefont {Serreze}}]{Stroeve:2007aa}%
  \BibitemOpen
  \bibfield  {author} {\bibinfo {author} {\bibfnamefont {J.}~\bibnamefont
  {Stroeve}}, \bibinfo {author} {\bibfnamefont {M.~M.}\ \bibnamefont
  {Holland}}, \bibinfo {author} {\bibfnamefont {W.}~\bibnamefont {Meier}},
  \bibinfo {author} {\bibfnamefont {T.}~\bibnamefont {Scambos}}, \ and\
  \bibinfo {author} {\bibfnamefont {M.}~\bibnamefont {Serreze}},\ }\href@noop
  {} {\bibfield  {journal} {\bibinfo  {journal} {Geophysical research letters}\
  }\textbf {\bibinfo {volume} {34}} (\bibinfo {year} {2007})}\BibitemShut
  {NoStop}%
\bibitem [{\citenamefont {Comiso}\ \emph {et~al.}(2008)\citenamefont {Comiso},
  \citenamefont {Parkinson}, \citenamefont {Gersten},\ and\ \citenamefont
  {Stock}}]{Comiso:2008aa}%
  \BibitemOpen
  \bibfield  {author} {\bibinfo {author} {\bibfnamefont {J.~C.}\ \bibnamefont
  {Comiso}}, \bibinfo {author} {\bibfnamefont {C.~L.}\ \bibnamefont
  {Parkinson}}, \bibinfo {author} {\bibfnamefont {R.}~\bibnamefont {Gersten}},
  \ and\ \bibinfo {author} {\bibfnamefont {L.}~\bibnamefont {Stock}},\
  }\href@noop {} {\bibfield  {journal} {\bibinfo  {journal} {Geophysical
  research letters}\ }\textbf {\bibinfo {volume} {35}} (\bibinfo {year}
  {2008})}\BibitemShut {NoStop}%
\bibitem [{\citenamefont {Kwok}\ and\ \citenamefont
  {Untersteiner}(2011)}]{OneWatt}%
  \BibitemOpen
  \bibfield  {author} {\bibinfo {author} {\bibfnamefont {R.}~\bibnamefont
  {Kwok}}\ and\ \bibinfo {author} {\bibfnamefont {N.}~\bibnamefont
  {Untersteiner}},\ }\href@noop {} {\bibfield  {journal} {\bibinfo  {journal}
  {Phys. Today}\ }\textbf {\bibinfo {volume} {64}},\ \bibinfo {pages} {36}
  (\bibinfo {year} {2011})}\BibitemShut {NoStop}%
\bibitem [{\citenamefont {Maykut}\ and\ \citenamefont
  {Untersteiner}(1971)}]{Maykut:1971aa}%
  \BibitemOpen
  \bibfield  {author} {\bibinfo {author} {\bibfnamefont {G.~A.}\ \bibnamefont
  {Maykut}}\ and\ \bibinfo {author} {\bibfnamefont {N.}~\bibnamefont
  {Untersteiner}},\ }\href {http://dx.doi.org/10.1029/JC076i006p01550}
  {\bibfield  {journal} {\bibinfo  {journal} {Journal of Geophysical Research}\
  }\textbf {\bibinfo {volume} {76}},\ \bibinfo {pages} {1550} (\bibinfo {year}
  {1971})}\BibitemShut {NoStop}%
\bibitem [{\citenamefont {Eisenman}\ and\ \citenamefont
  {Wettlaufer}(2009)}]{EW09}%
  \BibitemOpen
  \bibfield  {author} {\bibinfo {author} {\bibfnamefont {I.}~\bibnamefont
  {Eisenman}}\ and\ \bibinfo {author} {\bibfnamefont {J.~S.}\ \bibnamefont
  {Wettlaufer}},\ }\href@noop {} {\bibfield  {journal} {\bibinfo  {journal}
  {Proc. Natl. Acad. Sci. USA}\ }\textbf {\bibinfo {volume} {106}},\ \bibinfo
  {pages} {28} (\bibinfo {year} {2009})}\BibitemShut {NoStop}%
\bibitem [{\citenamefont {Tietsche}\ \emph {et~al.}(2011)\citenamefont
  {Tietsche}, \citenamefont {Notz}, \citenamefont {Jungclaus},\ and\
  \citenamefont {Marotzke}}]{Tietsche:2011}%
  \BibitemOpen
  \bibfield  {author} {\bibinfo {author} {\bibfnamefont {S.}~\bibnamefont
  {Tietsche}}, \bibinfo {author} {\bibfnamefont {D.}~\bibnamefont {Notz}},
  \bibinfo {author} {\bibfnamefont {J.~H.}\ \bibnamefont {Jungclaus}}, \ and\
  \bibinfo {author} {\bibfnamefont {J.}~\bibnamefont {Marotzke}},\ }\href@noop
  {} {\bibfield  {journal} {\bibinfo  {journal} {Geophys. Res. Lett.}\ }\textbf
  {\bibinfo {volume} {38}},\ \bibinfo {pages} {L02707} (\bibinfo {year}
  {2011})}\BibitemShut {NoStop}%
\bibitem [{\citenamefont {Serreze}\ and\ \citenamefont
  {Barry}(2011)}]{SerrezeBarry:2011}%
  \BibitemOpen
  \bibfield  {author} {\bibinfo {author} {\bibfnamefont {M.~C.}\ \bibnamefont
  {Serreze}}\ and\ \bibinfo {author} {\bibfnamefont {R.~G.}\ \bibnamefont
  {Barry}},\ }\href@noop {} {\bibfield  {journal} {\bibinfo  {journal} {Global
  and Planetary Change}\ }\textbf {\bibinfo {volume} {77}},\ \bibinfo {pages}
  {85} (\bibinfo {year} {2011})}\BibitemShut {NoStop}%
\bibitem [{\citenamefont {Moon}\ and\ \citenamefont
  {Wettlaufer}(2011)}]{MW:2011}%
  \BibitemOpen
  \bibfield  {author} {\bibinfo {author} {\bibfnamefont {W.}~\bibnamefont
  {Moon}}\ and\ \bibinfo {author} {\bibfnamefont {J.~S.}\ \bibnamefont
  {Wettlaufer}},\ }\href@noop {} {\bibfield  {journal} {\bibinfo  {journal}
  {Europhys. Lett.}\ }\textbf {\bibinfo {volume} {96}},\ \bibinfo {pages}
  {39001} (\bibinfo {year} {2011})}\BibitemShut {NoStop}%
\bibitem [{\citenamefont {Moon}\ and\ \citenamefont
  {Wettlaufer}(2012)}]{MW:2012}%
  \BibitemOpen
  \bibfield  {author} {\bibinfo {author} {\bibfnamefont {W.}~\bibnamefont
  {Moon}}\ and\ \bibinfo {author} {\bibfnamefont {J.~S.}\ \bibnamefont
  {Wettlaufer}},\ }\href@noop {} {\bibfield  {journal} {\bibinfo  {journal} {J.
  Geophys. Res.-Oceans}\ }\textbf {\bibinfo {volume} {117}},\ \bibinfo {pages}
  {C07007} (\bibinfo {year} {2012})}\BibitemShut {NoStop}%
\bibitem [{\citenamefont {Toppaladoddi}\ and\ \citenamefont
  {Wettlaufer}(2017)}]{Toppaladoddi:2017aa}%
  \BibitemOpen
  \bibfield  {author} {\bibinfo {author} {\bibfnamefont {S.}~\bibnamefont
  {Toppaladoddi}}\ and\ \bibinfo {author} {\bibfnamefont {J.~S.}\ \bibnamefont
  {Wettlaufer}},\ }\href {http://dx.doi.org/10.1007/s10955-016-1704-8}
  {\bibfield  {journal} {\bibinfo  {journal} {Journal of Statistical Physics}\
  ,\ \bibinfo {pages} {1}} (\bibinfo {year} {2017})}\BibitemShut {NoStop}%
\bibitem [{\citenamefont {Manley}(1995)}]{Manley:1995aa}%
  \BibitemOpen
  \bibfield  {author} {\bibinfo {author} {\bibfnamefont {T.}~\bibnamefont
  {Manley}},\ }\href@noop {} {\bibfield  {journal} {\bibinfo  {journal}
  {Journal of Geophysical Research: Oceans}\ }\textbf {\bibinfo {volume}
  {100}},\ \bibinfo {pages} {20627} (\bibinfo {year} {1995})}\BibitemShut
  {NoStop}%
\bibitem [{\citenamefont {Woodgate}\ and\ \citenamefont
  {Aagaard}(2005)}]{Woodgate:2005aa}%
  \BibitemOpen
  \bibfield  {author} {\bibinfo {author} {\bibfnamefont {R.~A.}\ \bibnamefont
  {Woodgate}}\ and\ \bibinfo {author} {\bibfnamefont {K.}~\bibnamefont
  {Aagaard}},\ }\href@noop {} {\bibfield  {journal} {\bibinfo  {journal}
  {Geophysical Research Letters}\ }\textbf {\bibinfo {volume} {32}} (\bibinfo
  {year} {2005})}\BibitemShut {NoStop}%
\bibitem [{\citenamefont {Kwok}\ \emph {et~al.}(2004)\citenamefont {Kwok},
  \citenamefont {Cunningham},\ and\ \citenamefont {Pang}}]{Kwok:2004aa}%
  \BibitemOpen
  \bibfield  {author} {\bibinfo {author} {\bibfnamefont {R.}~\bibnamefont
  {Kwok}}, \bibinfo {author} {\bibfnamefont {G.}~\bibnamefont {Cunningham}}, \
  and\ \bibinfo {author} {\bibfnamefont {S.}~\bibnamefont {Pang}},\ }\href@noop
  {} {\bibfield  {journal} {\bibinfo  {journal} {Journal of Geophysical
  Research: Oceans}\ }\textbf {\bibinfo {volume} {109}} (\bibinfo {year}
  {2004})}\BibitemShut {NoStop}%
\bibitem [{\citenamefont {Meredith}\ \emph {et~al.}(2001)\citenamefont
  {Meredith}, \citenamefont {Heywood}, \citenamefont {Dennis}, \citenamefont
  {Goldson}, \citenamefont {White}, \citenamefont {Fahrbach}, \citenamefont
  {Schauer},\ and\ \citenamefont {{\O}sterhus}}]{Meredith:2001aa}%
  \BibitemOpen
  \bibfield  {author} {\bibinfo {author} {\bibfnamefont {M.}~\bibnamefont
  {Meredith}}, \bibinfo {author} {\bibfnamefont {K.}~\bibnamefont {Heywood}},
  \bibinfo {author} {\bibfnamefont {P.}~\bibnamefont {Dennis}}, \bibinfo
  {author} {\bibfnamefont {L.}~\bibnamefont {Goldson}}, \bibinfo {author}
  {\bibfnamefont {R.}~\bibnamefont {White}}, \bibinfo {author} {\bibfnamefont
  {E.}~\bibnamefont {Fahrbach}}, \bibinfo {author} {\bibfnamefont
  {U.}~\bibnamefont {Schauer}}, \ and\ \bibinfo {author} {\bibfnamefont
  {S.}~\bibnamefont {{\O}sterhus}},\ }\href@noop {} {\bibfield  {journal}
  {\bibinfo  {journal} {Geophysical research letters}\ }\textbf {\bibinfo
  {volume} {28}},\ \bibinfo {pages} {1615} (\bibinfo {year}
  {2001})}\BibitemShut {NoStop}%
\bibitem [{\citenamefont {Untersteiner}(1988)}]{Untersteiner:1988aa}%
  \BibitemOpen
  \bibfield  {author} {\bibinfo {author} {\bibfnamefont {N.}~\bibnamefont
  {Untersteiner}},\ }\href {http://dx.doi.org/10.1029/JC093iC01p00527}
  {\bibfield  {journal} {\bibinfo  {journal} {Journal of Geophysical Research:
  Oceans}\ }\textbf {\bibinfo {volume} {93}},\ \bibinfo {pages} {527} (\bibinfo
  {year} {1988})}\BibitemShut {NoStop}%
\bibitem [{\citenamefont {Vinje}\ and\ \citenamefont
  {Finnek{\aa}sa}(1986)}]{Vinje:1986aa}%
  \BibitemOpen
  \bibfield  {author} {\bibinfo {author} {\bibfnamefont {T.}~\bibnamefont
  {Vinje}}\ and\ \bibinfo {author} {\bibfnamefont {{\O}.}~\bibnamefont
  {Finnek{\aa}sa}},\ }\href@noop {} {\emph {\bibinfo {title} {The ice transport
  through the Fram Strait}}}\ (\bibinfo  {publisher} {Norsk Polarinstitutt
  Oslo},\ \bibinfo {year} {1986})\BibitemShut {NoStop}%
\bibitem [{\citenamefont {Vinje}\ \emph {et~al.}(1998)\citenamefont {Vinje},
  \citenamefont {Nordlund},\ and\ \citenamefont {Kvambekk}}]{Vinje:1998aa}%
  \BibitemOpen
  \bibfield  {author} {\bibinfo {author} {\bibfnamefont {T.}~\bibnamefont
  {Vinje}}, \bibinfo {author} {\bibfnamefont {N.}~\bibnamefont {Nordlund}}, \
  and\ \bibinfo {author} {\bibfnamefont {{\AA}.}~\bibnamefont {Kvambekk}},\
  }\href@noop {} {\bibfield  {journal} {\bibinfo  {journal} {Journal of
  Geophysical Research: Oceans}\ }\textbf {\bibinfo {volume} {103}},\ \bibinfo
  {pages} {10437} (\bibinfo {year} {1998})}\BibitemShut {NoStop}%
\bibitem [{\citenamefont {Aagaard}\ and\ \citenamefont
  {Coachman}(1968{\natexlab{a}})}]{Aagaard:1968aa}%
  \BibitemOpen
  \bibfield  {author} {\bibinfo {author} {\bibfnamefont {K.}~\bibnamefont
  {Aagaard}}\ and\ \bibinfo {author} {\bibfnamefont {L.~K.}\ \bibnamefont
  {Coachman}},\ }\href@noop {} {\bibfield  {journal} {\bibinfo  {journal}
  {Arctic}\ ,\ \bibinfo {pages} {181}} (\bibinfo {year}
  {1968}{\natexlab{a}})}\BibitemShut {NoStop}%
\bibitem [{\citenamefont {Aagaard}\ and\ \citenamefont
  {Coachman}(1968{\natexlab{b}})}]{Aagaard:1968ab}%
  \BibitemOpen
  \bibfield  {author} {\bibinfo {author} {\bibfnamefont {K.}~\bibnamefont
  {Aagaard}}\ and\ \bibinfo {author} {\bibfnamefont {L.}~\bibnamefont
  {Coachman}},\ }\href@noop {} {\bibfield  {journal} {\bibinfo  {journal}
  {Arctic}\ ,\ \bibinfo {pages} {267}} (\bibinfo {year}
  {1968}{\natexlab{b}})}\BibitemShut {NoStop}%
\bibitem [{\citenamefont {Aagaard}\ and\ \citenamefont
  {Carmack}(1989)}]{Aagaard:1989aa}%
  \BibitemOpen
  \bibfield  {author} {\bibinfo {author} {\bibfnamefont {K.}~\bibnamefont
  {Aagaard}}\ and\ \bibinfo {author} {\bibfnamefont {E.~C.}\ \bibnamefont
  {Carmack}},\ }\href@noop {} {\bibfield  {journal} {\bibinfo  {journal}
  {Journal of Geophysical Research: Oceans}\ }\textbf {\bibinfo {volume}
  {94}},\ \bibinfo {pages} {14485} (\bibinfo {year} {1989})}\BibitemShut
  {NoStop}%
\bibitem [{\citenamefont {Rudels}\ and\ \citenamefont
  {Quadfasel}(1991)}]{Rudels:1991aa}%
  \BibitemOpen
  \bibfield  {author} {\bibinfo {author} {\bibfnamefont {B.}~\bibnamefont
  {Rudels}}\ and\ \bibinfo {author} {\bibfnamefont {D.}~\bibnamefont
  {Quadfasel}},\ }\href@noop {} {\bibfield  {journal} {\bibinfo  {journal}
  {Journal of Marine Systems}\ }\textbf {\bibinfo {volume} {2}},\ \bibinfo
  {pages} {435} (\bibinfo {year} {1991})}\BibitemShut {NoStop}%
\bibitem [{\citenamefont {de~Steur}\ \emph {et~al.}(2009)\citenamefont
  {de~Steur}, \citenamefont {Hansen}, \citenamefont {Gerdes}, \citenamefont
  {Karcher}, \citenamefont {Fahrbach},\ and\ \citenamefont
  {Holfort}}]{Steur:2009aa}%
  \BibitemOpen
  \bibfield  {author} {\bibinfo {author} {\bibfnamefont {L.}~\bibnamefont
  {de~Steur}}, \bibinfo {author} {\bibfnamefont {E.}~\bibnamefont {Hansen}},
  \bibinfo {author} {\bibfnamefont {R.}~\bibnamefont {Gerdes}}, \bibinfo
  {author} {\bibfnamefont {M.}~\bibnamefont {Karcher}}, \bibinfo {author}
  {\bibfnamefont {E.}~\bibnamefont {Fahrbach}}, \ and\ \bibinfo {author}
  {\bibfnamefont {J.}~\bibnamefont {Holfort}},\ }\href@noop {} {\bibfield
  {journal} {\bibinfo  {journal} {Geophysical Research Letters}\ }\textbf
  {\bibinfo {volume} {36}} (\bibinfo {year} {2009})}\BibitemShut {NoStop}%
\bibitem [{\citenamefont {Smedsrud}\ \emph {et~al.}(2017)\citenamefont
  {Smedsrud}, \citenamefont {Halvorsen}, \citenamefont {Stroeve}, \citenamefont
  {Zhang},\ and\ \citenamefont {Kloster}}]{Smedsrud:2017aa}%
  \BibitemOpen
  \bibfield  {author} {\bibinfo {author} {\bibfnamefont {L.~H.}\ \bibnamefont
  {Smedsrud}}, \bibinfo {author} {\bibfnamefont {M.~H.}\ \bibnamefont
  {Halvorsen}}, \bibinfo {author} {\bibfnamefont {J.~C.}\ \bibnamefont
  {Stroeve}}, \bibinfo {author} {\bibfnamefont {R.}~\bibnamefont {Zhang}}, \
  and\ \bibinfo {author} {\bibfnamefont {K.}~\bibnamefont {Kloster}},\
  }\href@noop {} {\bibfield  {journal} {\bibinfo  {journal} {The Cryosphere}\
  }\textbf {\bibinfo {volume} {11}},\ \bibinfo {pages} {65} (\bibinfo {year}
  {2017})}\BibitemShut {NoStop}%
\bibitem [{\citenamefont {Colony}\ and\ \citenamefont
  {Thorndike}(1985)}]{Colony:1985}%
  \BibitemOpen
  \bibfield  {author} {\bibinfo {author} {\bibfnamefont {R.}~\bibnamefont
  {Colony}}\ and\ \bibinfo {author} {\bibfnamefont {A.~S.}\ \bibnamefont
  {Thorndike}},\ }\href@noop {} {\bibfield  {journal} {\bibinfo  {journal}
  {Journal of Geophysical Research: Oceans}\ }\textbf {\bibinfo {volume}
  {90}},\ \bibinfo {pages} {965} (\bibinfo {year} {1985})}\BibitemShut
  {NoStop}%
\bibitem [{\citenamefont {Mauritzen}\ and\ \citenamefont
  {H{\"a}kkinen}(1997)}]{Mauritzen:1997aa}%
  \BibitemOpen
  \bibfield  {author} {\bibinfo {author} {\bibfnamefont {C.}~\bibnamefont
  {Mauritzen}}\ and\ \bibinfo {author} {\bibfnamefont {S.}~\bibnamefont
  {H{\"a}kkinen}},\ }\href@noop {} {\bibfield  {journal} {\bibinfo  {journal}
  {Geophysical Research Letters}\ }\textbf {\bibinfo {volume} {24}},\ \bibinfo
  {pages} {3257} (\bibinfo {year} {1997})}\BibitemShut {NoStop}%
\bibitem [{\citenamefont {Jungclaus}\ \emph {et~al.}(2005)\citenamefont
  {Jungclaus}, \citenamefont {Haak}, \citenamefont {Latif},\ and\ \citenamefont
  {Mikolajewicz}}]{Jungclaus:2005aa}%
  \BibitemOpen
  \bibfield  {author} {\bibinfo {author} {\bibfnamefont {J.~H.}\ \bibnamefont
  {Jungclaus}}, \bibinfo {author} {\bibfnamefont {H.}~\bibnamefont {Haak}},
  \bibinfo {author} {\bibfnamefont {M.}~\bibnamefont {Latif}}, \ and\ \bibinfo
  {author} {\bibfnamefont {U.}~\bibnamefont {Mikolajewicz}},\ }\href@noop {}
  {\bibfield  {journal} {\bibinfo  {journal} {Journal of climate}\ }\textbf
  {\bibinfo {volume} {18}},\ \bibinfo {pages} {4013} (\bibinfo {year}
  {2005})}\BibitemShut {NoStop}%
\bibitem [{\citenamefont {Koenigk}\ \emph {et~al.}(2006)\citenamefont
  {Koenigk}, \citenamefont {Mikolajewicz}, \citenamefont {Haak},\ and\
  \citenamefont {Jungclaus}}]{Koenigk:2006aa}%
  \BibitemOpen
  \bibfield  {author} {\bibinfo {author} {\bibfnamefont {T.}~\bibnamefont
  {Koenigk}}, \bibinfo {author} {\bibfnamefont {U.}~\bibnamefont
  {Mikolajewicz}}, \bibinfo {author} {\bibfnamefont {H.}~\bibnamefont {Haak}},
  \ and\ \bibinfo {author} {\bibfnamefont {J.}~\bibnamefont {Jungclaus}},\
  }\href@noop {} {\bibfield  {journal} {\bibinfo  {journal} {Climate dynamics}\
  }\textbf {\bibinfo {volume} {26}},\ \bibinfo {pages} {17} (\bibinfo {year}
  {2006})}\BibitemShut {NoStop}%
\bibitem [{\citenamefont {Miles}\ \emph {et~al.}(2014)\citenamefont {Miles},
  \citenamefont {Divine}, \citenamefont {Furevik}, \citenamefont {Jansen},
  \citenamefont {Moros},\ and\ \citenamefont {Ogilvie}}]{Miles:2014aa}%
  \BibitemOpen
  \bibfield  {author} {\bibinfo {author} {\bibfnamefont {M.~W.}\ \bibnamefont
  {Miles}}, \bibinfo {author} {\bibfnamefont {D.~V.}\ \bibnamefont {Divine}},
  \bibinfo {author} {\bibfnamefont {T.}~\bibnamefont {Furevik}}, \bibinfo
  {author} {\bibfnamefont {E.}~\bibnamefont {Jansen}}, \bibinfo {author}
  {\bibfnamefont {M.}~\bibnamefont {Moros}}, \ and\ \bibinfo {author}
  {\bibfnamefont {A.~E.}\ \bibnamefont {Ogilvie}},\ }\href@noop {} {\bibfield
  {journal} {\bibinfo  {journal} {Geophysical Research Letters}\ }\textbf
  {\bibinfo {volume} {41}},\ \bibinfo {pages} {463} (\bibinfo {year}
  {2014})}\BibitemShut {NoStop}%
\bibitem [{\citenamefont {Vowinckel}\ and\ \citenamefont
  {Orvig}(1962)}]{Vowinckel:1962aa}%
  \BibitemOpen
  \bibfield  {author} {\bibinfo {author} {\bibfnamefont {E.}~\bibnamefont
  {Vowinckel}}\ and\ \bibinfo {author} {\bibfnamefont {S.}~\bibnamefont
  {Orvig}},\ }\href@noop {} {\bibfield  {journal} {\bibinfo  {journal}
  {Arctic}\ }\textbf {\bibinfo {volume} {15}},\ \bibinfo {pages} {205}
  (\bibinfo {year} {1962})}\BibitemShut {NoStop}%
\bibitem [{\citenamefont {Aagaard}\ \emph {et~al.}(1987)\citenamefont
  {Aagaard}, \citenamefont {Foldvik},\ and\ \citenamefont
  {Hillman}}]{Aagaard:1987aa}%
  \BibitemOpen
  \bibfield  {author} {\bibinfo {author} {\bibfnamefont {K.}~\bibnamefont
  {Aagaard}}, \bibinfo {author} {\bibfnamefont {A.}~\bibnamefont {Foldvik}}, \
  and\ \bibinfo {author} {\bibfnamefont {S.}~\bibnamefont {Hillman}},\
  }\href@noop {} {\bibfield  {journal} {\bibinfo  {journal} {Journal of
  Geophysical Research: Oceans}\ }\textbf {\bibinfo {volume} {92}},\ \bibinfo
  {pages} {3778} (\bibinfo {year} {1987})}\BibitemShut {NoStop}%
\bibitem [{\citenamefont {McPhee}(2008)}]{McPhee:2008aa}%
  \BibitemOpen
  \bibfield  {author} {\bibinfo {author} {\bibfnamefont {M.}~\bibnamefont
  {McPhee}},\ }\href@noop {} {\emph {\bibinfo {title} {Air-ice-ocean
  interaction: Turbulent ocean boundary layer exchange processes}}}\ (\bibinfo
  {publisher} {Springer Science \& Business Media},\ \bibinfo {year}
  {2008})\BibitemShut {NoStop}%
\bibitem [{\citenamefont {Omstedt}\ and\ \citenamefont
  {Wettlaufer}(1992)}]{Omstedt:1992aa}%
  \BibitemOpen
  \bibfield  {author} {\bibinfo {author} {\bibfnamefont {A.}~\bibnamefont
  {Omstedt}}\ and\ \bibinfo {author} {\bibfnamefont {J.}~\bibnamefont
  {Wettlaufer}},\ }\href@noop {} {\bibfield  {journal} {\bibinfo  {journal}
  {Journal of Geophysical Research: Oceans}\ }\textbf {\bibinfo {volume}
  {97}},\ \bibinfo {pages} {9383} (\bibinfo {year} {1992})}\BibitemShut
  {NoStop}%
\bibitem [{\citenamefont {Foldvik}\ \emph {et~al.}(1988)\citenamefont
  {Foldvik}, \citenamefont {Aagaard},\ and\ \citenamefont
  {T{\o}rresen}}]{Foldvik:1988aa}%
  \BibitemOpen
  \bibfield  {author} {\bibinfo {author} {\bibfnamefont {A.}~\bibnamefont
  {Foldvik}}, \bibinfo {author} {\bibfnamefont {K.}~\bibnamefont {Aagaard}}, \
  and\ \bibinfo {author} {\bibfnamefont {T.}~\bibnamefont {T{\o}rresen}},\
  }\href@noop {} {\bibfield  {journal} {\bibinfo  {journal} {Deep Sea Research
  Part A. Oceanographic Research Papers}\ }\textbf {\bibinfo {volume} {35}},\
  \bibinfo {pages} {1335} (\bibinfo {year} {1988})}\BibitemShut {NoStop}%
\bibitem [{\citenamefont {Woodgate}\ \emph {et~al.}(1999)\citenamefont
  {Woodgate}, \citenamefont {Fahrbach},\ and\ \citenamefont
  {Rohardt}}]{Woodgate:1999aa}%
  \BibitemOpen
  \bibfield  {author} {\bibinfo {author} {\bibfnamefont {R.~A.}\ \bibnamefont
  {Woodgate}}, \bibinfo {author} {\bibfnamefont {E.}~\bibnamefont {Fahrbach}},
  \ and\ \bibinfo {author} {\bibfnamefont {G.}~\bibnamefont {Rohardt}},\
  }\href@noop {} {\bibfield  {journal} {\bibinfo  {journal} {Journal of
  Geophysical Research: Oceans}\ }\textbf {\bibinfo {volume} {104}},\ \bibinfo
  {pages} {18059} (\bibinfo {year} {1999})}\BibitemShut {NoStop}%
\bibitem [{\citenamefont {Cokelet}\ \emph {et~al.}(2008)\citenamefont
  {Cokelet}, \citenamefont {Tervalon},\ and\ \citenamefont
  {Bellingham}}]{Cokelet:2008aa}%
  \BibitemOpen
  \bibfield  {author} {\bibinfo {author} {\bibfnamefont {E.~D.}\ \bibnamefont
  {Cokelet}}, \bibinfo {author} {\bibfnamefont {N.}~\bibnamefont {Tervalon}}, \
  and\ \bibinfo {author} {\bibfnamefont {J.~G.}\ \bibnamefont {Bellingham}},\
  }\href@noop {} {\bibfield  {journal} {\bibinfo  {journal} {Journal of
  Geophysical Research: Oceans}\ }\textbf {\bibinfo {volume} {113}} (\bibinfo
  {year} {2008})}\BibitemShut {NoStop}%
\bibitem [{\citenamefont {Sutherland}\ and\ \citenamefont
  {Pickart}(2008)}]{Sutherland:2008aa}%
  \BibitemOpen
  \bibfield  {author} {\bibinfo {author} {\bibfnamefont {D.~A.}\ \bibnamefont
  {Sutherland}}\ and\ \bibinfo {author} {\bibfnamefont {R.~S.}\ \bibnamefont
  {Pickart}},\ }\href@noop {} {\bibfield  {journal} {\bibinfo  {journal}
  {Progress in Oceanography}\ }\textbf {\bibinfo {volume} {78}},\ \bibinfo
  {pages} {58} (\bibinfo {year} {2008})}\BibitemShut {NoStop}%
\bibitem [{\citenamefont {Boyd}\ and\ \citenamefont
  {D'Asaro}(1994)}]{Boyd:1994aa}%
  \BibitemOpen
  \bibfield  {author} {\bibinfo {author} {\bibfnamefont {T.~J.}\ \bibnamefont
  {Boyd}}\ and\ \bibinfo {author} {\bibfnamefont {E.~A.}\ \bibnamefont
  {D'Asaro}},\ }\href@noop {} {\bibfield  {journal} {\bibinfo  {journal}
  {Journal of Geophysical Research: Oceans}\ }\textbf {\bibinfo {volume}
  {99}},\ \bibinfo {pages} {22597} (\bibinfo {year} {1994})}\BibitemShut
  {NoStop}%
\bibitem [{\citenamefont {Rohde}\ \emph {et~al.}(2013)\citenamefont {Rohde},
  \citenamefont {Muller}, \citenamefont {Jacobsen}, \citenamefont {Muller},
  \citenamefont {Perlmutter}, \citenamefont {Rosenfeld}, \citenamefont
  {Wurtele}, \citenamefont {Groom},\ and\ \citenamefont
  {Wickham}}]{Rohde:2013aa}%
  \BibitemOpen
  \bibfield  {author} {\bibinfo {author} {\bibfnamefont {R.}~\bibnamefont
  {Rohde}}, \bibinfo {author} {\bibfnamefont {R.}~\bibnamefont {Muller}},
  \bibinfo {author} {\bibfnamefont {R.}~\bibnamefont {Jacobsen}}, \bibinfo
  {author} {\bibfnamefont {E.}~\bibnamefont {Muller}}, \bibinfo {author}
  {\bibfnamefont {S.}~\bibnamefont {Perlmutter}}, \bibinfo {author}
  {\bibfnamefont {A.}~\bibnamefont {Rosenfeld}}, \bibinfo {author}
  {\bibfnamefont {J.}~\bibnamefont {Wurtele}}, \bibinfo {author} {\bibfnamefont
  {D.}~\bibnamefont {Groom}}, \ and\ \bibinfo {author} {\bibfnamefont
  {C.}~\bibnamefont {Wickham}},\ }\href@noop {} {\bibfield  {journal} {\bibinfo
   {journal} {of}\ }\textbf {\bibinfo {volume} {7}},\ \bibinfo {pages} {2}
  (\bibinfo {year} {2013})}\BibitemShut {NoStop}%
\bibitem [{\citenamefont {Agarwal}\ and\ \citenamefont
  {Wettlaufer}(2017)}]{Agarwal:2017aa}%
  \BibitemOpen
  \bibfield  {author} {\bibinfo {author} {\bibfnamefont {S.}~\bibnamefont
  {Agarwal}}\ and\ \bibinfo {author} {\bibfnamefont {J.~S.}\ \bibnamefont
  {Wettlaufer}},\ }\bibfield  {booktitle} {\emph {\bibinfo {booktitle} {Journal
  of Climate}},\ }\href@noop {} {\bibfield  {journal} {\bibinfo  {journal}
  {Journal of Climate}\ }\textbf {\bibinfo {volume} {30}},\ \bibinfo {pages}
  {4873} (\bibinfo {year} {2017})}\BibitemShut {NoStop}%
\bibitem [{\citenamefont {Timmermann}\ \emph {et~al.}(2003)\citenamefont
  {Timmermann}, \citenamefont {Gildor}, \citenamefont {Schulz},\ and\
  \citenamefont {Tziperman}}]{Timmermann:2003aa}%
  \BibitemOpen
  \bibfield  {author} {\bibinfo {author} {\bibfnamefont {A.}~\bibnamefont
  {Timmermann}}, \bibinfo {author} {\bibfnamefont {H.}~\bibnamefont {Gildor}},
  \bibinfo {author} {\bibfnamefont {M.}~\bibnamefont {Schulz}}, \ and\ \bibinfo
  {author} {\bibfnamefont {E.}~\bibnamefont {Tziperman}},\ }\bibfield
  {booktitle} {\emph {\bibinfo {booktitle} {Journal of Climate}},\ }\href
  {https://doi.org/10.1175/1520-0442(2003)016<2569:CRMCOT>2.0.CO;2} {\bibfield
  {journal} {\bibinfo  {journal} {Journal of Climate}\ }\textbf {\bibinfo
  {volume} {16}},\ \bibinfo {pages} {2569} (\bibinfo {year}
  {2003})}\BibitemShut {NoStop}%
\bibitem [{\citenamefont {Sevellec}\ \emph {et~al.}(2017)\citenamefont
  {Sevellec}, \citenamefont {Fedorov},\ and\ \citenamefont
  {Liu}}]{Sevellec:2017aa}%
  \BibitemOpen
  \bibfield  {author} {\bibinfo {author} {\bibfnamefont {F.}~\bibnamefont
  {Sevellec}}, \bibinfo {author} {\bibfnamefont {A.~V.}\ \bibnamefont
  {Fedorov}}, \ and\ \bibinfo {author} {\bibfnamefont {W.}~\bibnamefont
  {Liu}},\ }\href {http://dx.doi.org/10.1038/nclimate3353} {\bibfield
  {journal} {\bibinfo  {journal} {Nature Clim. Change}\ }\textbf {\bibinfo
  {volume} {7}},\ \bibinfo {pages} {604} (\bibinfo {year} {2017})}\BibitemShut
  {NoStop}%
\bibitem [{\citenamefont {Walczowski}\ and\ \citenamefont
  {Piechura}(2007)}]{Walczowski:2007aa}%
  \BibitemOpen
  \bibfield  {author} {\bibinfo {author} {\bibfnamefont {W.}~\bibnamefont
  {Walczowski}}\ and\ \bibinfo {author} {\bibfnamefont {J.}~\bibnamefont
  {Piechura}},\ }\href {http://dx.doi.org/10.1029/2007GL029974} {\bibfield
  {journal} {\bibinfo  {journal} {Geophysical Research Letters}\ }\textbf
  {\bibinfo {volume} {34}},\ \bibinfo {pages} {n/a} (\bibinfo {year}
  {2007})}\BibitemShut {NoStop}%
\bibitem [{\citenamefont {Schauer}\ \emph {et~al.}(2008)\citenamefont
  {Schauer}, \citenamefont {Beszczynska-M{\"o}ller}, \citenamefont
  {Walczowski}, \citenamefont {Fahrbach}, \citenamefont {Piechura},\ and\
  \citenamefont {Hansen}}]{Schauer:2008aa}%
  \BibitemOpen
  \bibfield  {author} {\bibinfo {author} {\bibfnamefont {U.}~\bibnamefont
  {Schauer}}, \bibinfo {author} {\bibfnamefont {A.}~\bibnamefont
  {Beszczynska-M{\"o}ller}}, \bibinfo {author} {\bibfnamefont {W.}~\bibnamefont
  {Walczowski}}, \bibinfo {author} {\bibfnamefont {E.}~\bibnamefont
  {Fahrbach}}, \bibinfo {author} {\bibfnamefont {J.}~\bibnamefont {Piechura}},
  \ and\ \bibinfo {author} {\bibfnamefont {E.}~\bibnamefont {Hansen}},\
  }\enquote {\bibinfo {title} {Variation of measured heat flow through the fram
  strait between 1997 and 2006},}\ in\ \href
  {http://dx.doi.org/10.1007/978-1-4020-6774-7_4} {\emph {\bibinfo {booktitle}
  {Arctic--Subarctic Ocean Fluxes: Defining the Role of the Northern Seas in
  Climate}}}\ (\bibinfo  {publisher} {Springer Netherlands},\ \bibinfo
  {address} {Dordrecht},\ \bibinfo {year} {2008})\ pp.\ \bibinfo {pages}
  {65--85}\BibitemShut {NoStop}%
\bibitem [{\citenamefont {Thorndike}\ and\ \citenamefont
  {Colony}(1982)}]{Thorndike:1982}%
  \BibitemOpen
  \bibfield  {author} {\bibinfo {author} {\bibfnamefont {A.~S.}\ \bibnamefont
  {Thorndike}}\ and\ \bibinfo {author} {\bibfnamefont {R.}~\bibnamefont
  {Colony}},\ }\href@noop {} {\bibfield  {journal} {\bibinfo  {journal} {J.
  Geophys. Res.-Oceans}\ }\textbf {\bibinfo {volume} {87}},\ \bibinfo {pages}
  {5845} (\bibinfo {year} {1982})}\BibitemShut {NoStop}%
\end{thebibliography}
%

\newpage

\section*{Supplementary Information (SI)}

\subsection*{Additional Supporting Information (Files uploaded separately)}

\begin{enumerate}
\item Caption for Movie S1 : This simulation shows the seasonally varying daily sea-ice edge after the transient has been removed for two seasonal cycles. Not only is there an asymmetry between the growth and melt rates of sea-ice edge but also an asymmetry between when different velocity profiles reach their maximum extent.
\item Caption for Movie S2 : This simulation shows the daily flow of sea ice across the Fram Strait, with the thickness variation shown in colors. The sea-ice edge here corresponds to that in Movie S1.
\end{enumerate}

\subsection*{$\mathcal{F} = 1$, $\mathcal{B} = 1$}
Eqns. 1-4 in the main text forms the complete system to be solved to determine sea-ice thickness $h$, mixed-layer depth $H$, mixed-layer temperature $T_{ml}$ and mixed-layer salinity $S_{ml}$. In steady state, this coupled system can be written as

\begin{eqnarray}
F_b - F_a &=& -\rho L U_i\frac{\partial}{\partial x}  h - \rho c_p (T_{ml} - T_w) U_w\frac{\partial}{\partial x} H \nonumber \\
&&- \rho c_p H U_w\frac{\partial}{\partial x}T_{ml} \nonumber \\
F_b - F_{ml} &=& - \rho c_p (T_{ml} - T_w) U_w\frac{\partial}{\partial x} H - \rho c_p H U_w\frac{\partial}{\partial x}T_{ml} \nonumber  \\
-U_i S_w\frac{\partial}{\partial x}h &=& U_w (S_{ml} - S_w) \frac{\partial}{\partial x} H + U_w H \frac{\partial}{\partial x} (S_{ml} - S_w) \nonumber  \\
\alpha \frac{\partial}{\partial x} T_{ml} &=& \beta \frac{\partial}{\partial x} S_{ml} 
\label{eq:SS_system}
\end{eqnarray}

which gives,

\begin{eqnarray}
H (\mathcal{B} - 1)\frac{\partial}{\partial x} T_{ml} &=& \frac{\beta S_w(F_a - F_{ml}) [\mathcal{F} - \mathcal{B}]}{\alpha L\rho U_w} \nonumber \\
(S_w - S_{ml})(\mathcal{B} - 1)\frac{\partial}{\partial x} H &=& \frac{S_w(F_a - F_{ml}) [\mathcal{F} - 1]}{L\rho U_w}\nonumber \\
\frac{\partial}{\partial x} h &=& \frac{F_a - F_{ml}}{L \rho U_i} \nonumber \\
H (\mathcal{B} - 1)\frac{\partial}{\partial x} S_{ml} &=& \frac{S_w(F_a - F_{ml}) [\mathcal{F} - \mathcal{B}]}{L\rho U_w}
\label{eq:solve}
\end{eqnarray}
where, $\mathcal{B} = \alpha(T_w - T_{ml})/\beta(S_w - S_{ml})$, and $\mathcal{F} = \alpha L (F_b - F_{ml})/\beta c_p S_w (F_a - F_{ml})$. 
\\
From Eqn. \ref{eq:SS_system} and using $L'H\hat{o}pital$'s rule
\begin{eqnarray}
\lim_{x \rightarrow x_e}\mathcal{B} &=& \lim_{T_{ml} \rightarrow T_w, S_{ml} \rightarrow S_w}\mathcal{B} \nonumber \\
&=& \lim_{T_{ml} \rightarrow T_w, S_{ml} \rightarrow S_w} \frac{\alpha(T_w - T_{ml})}{\beta(S_w - S_{ml})} \nonumber \\
&=& \lim_{T_{ml} \rightarrow T_w, S_{ml} \rightarrow S_w} \frac{\alpha \frac{\partial}{\partial x} T_{ml}}{\beta \frac{\partial}{\partial x} S_{ml}} \nonumber \\
&=& 1
\end{eqnarray}

Now, let $\mathcal{B} = f(x)$
\begin{eqnarray}
\Rightarrow 
-\alpha \frac{\partial}{\partial x} T_{ml} &=& -f(x)\alpha \frac{\partial}{\partial x} S_{ml} + f'(x)\beta(S_w - S_{ml}) \nonumber \\
\Rightarrow -\beta \frac{\partial}{\partial x} S_{ml} &=& -f(x)\alpha \frac{\partial}{\partial x} S_{ml} + f'(x)\beta(S_w - S_{ml}) \nonumber \\
\Rightarrow (f(x) - 1)\frac{\partial}{\partial x} S_{ml} &=& f'(x)(S_w - S_{ml})
\label{eq:blow}
\end{eqnarray}
Therefore, $S_{ml}$ has a singularity at the ice edge unless $f(x) \equiv 1$ and $f'(x) \equiv 0$. Therefore,

\begin{equation}
\mathcal{B} \equiv 1.
\end{equation}

We can estimate the behavior of Eqns. \ref{eq:solve} near the ice edge as $H \rightarrow 0$ as follows:

\begin{eqnarray}
\lim_{H \rightarrow 0} \frac{S_w(F_a - F_{ml}) [\mathcal{F} - \mathcal{B}]}{L\rho U_w} = 0 \nonumber \\
\Rightarrow \lim_{x \rightarrow x_e} \alpha L(F_b - F_{ml}) - \beta Bc_pS_w(F_a - F_{ml}) = 0 \nonumber \\
\Rightarrow 
\lim_{x \rightarrow x_e}\mathcal{F}  = \lim_{x \rightarrow x_e}\mathcal{B} = 1.
\end{eqnarray}

At the ice edge, the system has a singular behavior since $H \rightarrow 0$. To overcome this singularity, we rearrange this system to solve for ice thickness $h$ and mixed-layer heat content $H(T_w - T_{ml})$. The coupled system of Equations \ref{eq:solve} can be written as:

\begin{eqnarray}
\frac{\partial}{\partial x}H(T_w - T_{ml}) &=& \frac{\beta S_w(F_a - F_{ml}) }{\alpha L\rho U_w} \nonumber \\
\frac{\partial}{\partial x} h &=& \frac{F_a - F_{ml}}{L \rho U_i} \nonumber \\
\mathcal{F} = \frac{\alpha L (F_b - F_{ml})}{\beta c_p S_w (F_a - F_{ml})} &=& 1 \nonumber \\
\mathcal{B} = \frac{\alpha (T_w - T_{ml})}{\beta (S_w - S_{ml})} &=& 1
\label{eq:solve2}
\end{eqnarray}
which are Eqns. 6 in the main text.
\eject
\begin{table}[]
\centering
\caption{Description of model parameters}
\label{tab:param}
\begin{tabular}{@{}ll@{}}
\hline
Symbol         & Description                       \\ \hline
$h$            & Sea-ice thickness                 \\
$H$            & Mixed-layer depth                 \\
$T_{ml}$       & Mixed-layer temperature           \\
$S_{ml}$       & Mixed-layer salinity              \\
$U_i$          & Sea-ice velocity                  \\
$U_w$          & Mixed-layer velocity              \\
$T_w$          & Open ocean temperature            \\
$S_w$          & Open ocean salinity               \\
$\lambda_{ml}$ & Heat transfer coefficient         \\
$St$           & Stanton number                    \\
$T_f$          & Liquidus temperature              \\
$T_s$          & Sea-ice surface temperature       \\
$T_a$          & Atmospheric temperature           \\
$F_{ml}$       & Heat flux to the underside of ice \\
$F_b$          & Abyssal heat flux                 \\
$F_a$          & Atmospheric heat flux             \\
$\mathcal{B}$  & Density ratio between the mixed layer and the ocean underneath                    \\
$\mathcal{F}$  & Energy conservation parameter              \\ \hline
\end{tabular}
\end{table}
\setcounter{figure}{0} 
\section*{Figures}
\renewcommand{\thefigure}{S\arabic{figure}}
\begin{figure}[htbp!]
        (a)\includegraphics[trim = 0 0 0 0, clip, width = 0.21\textwidth]{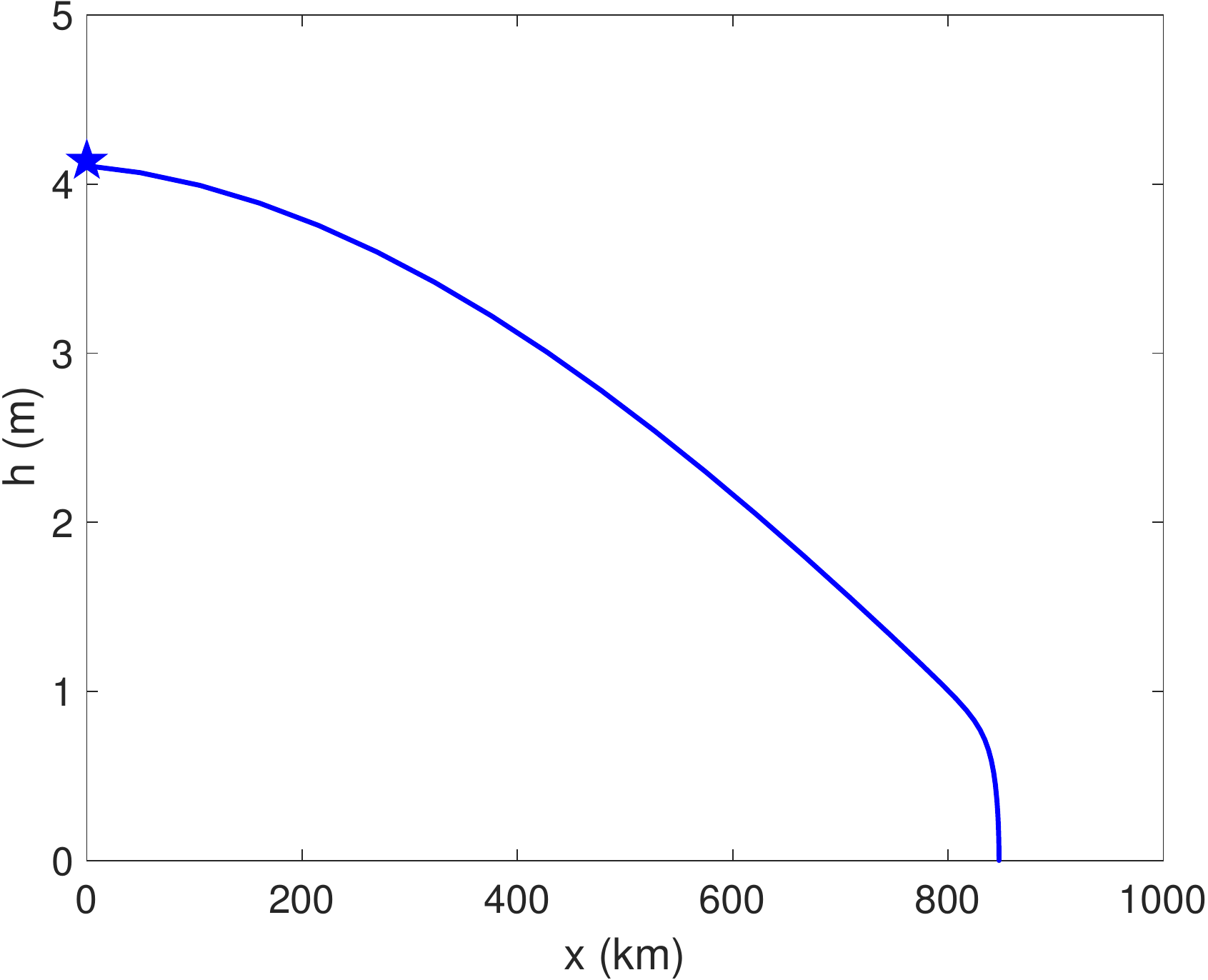}	  	
        (b)\includegraphics[trim = 0 0 0 0, clip, width = 0.21\textwidth]{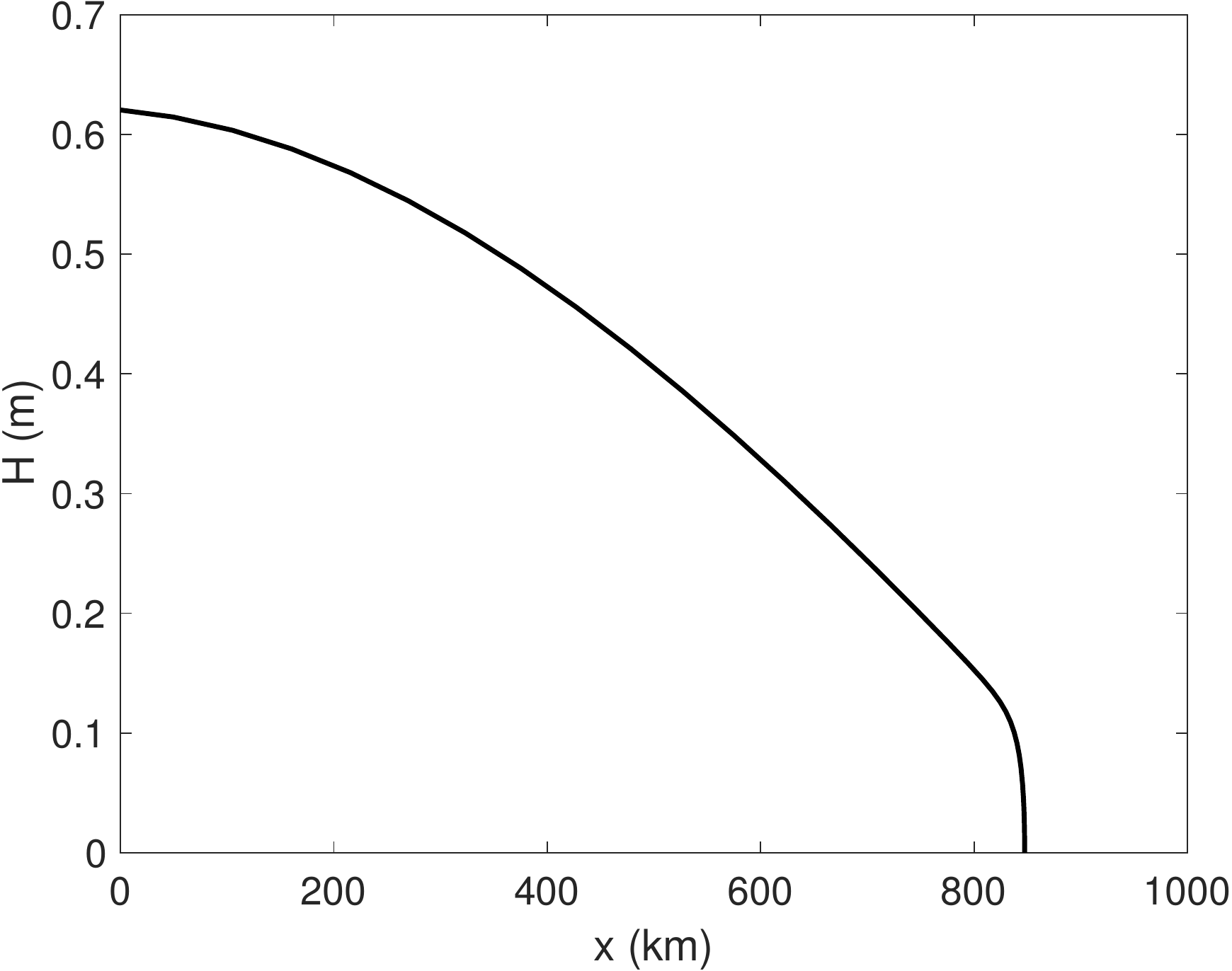}
        	
        (c)\includegraphics[trim = 0 0 0 0, clip, width = 0.21\textwidth]{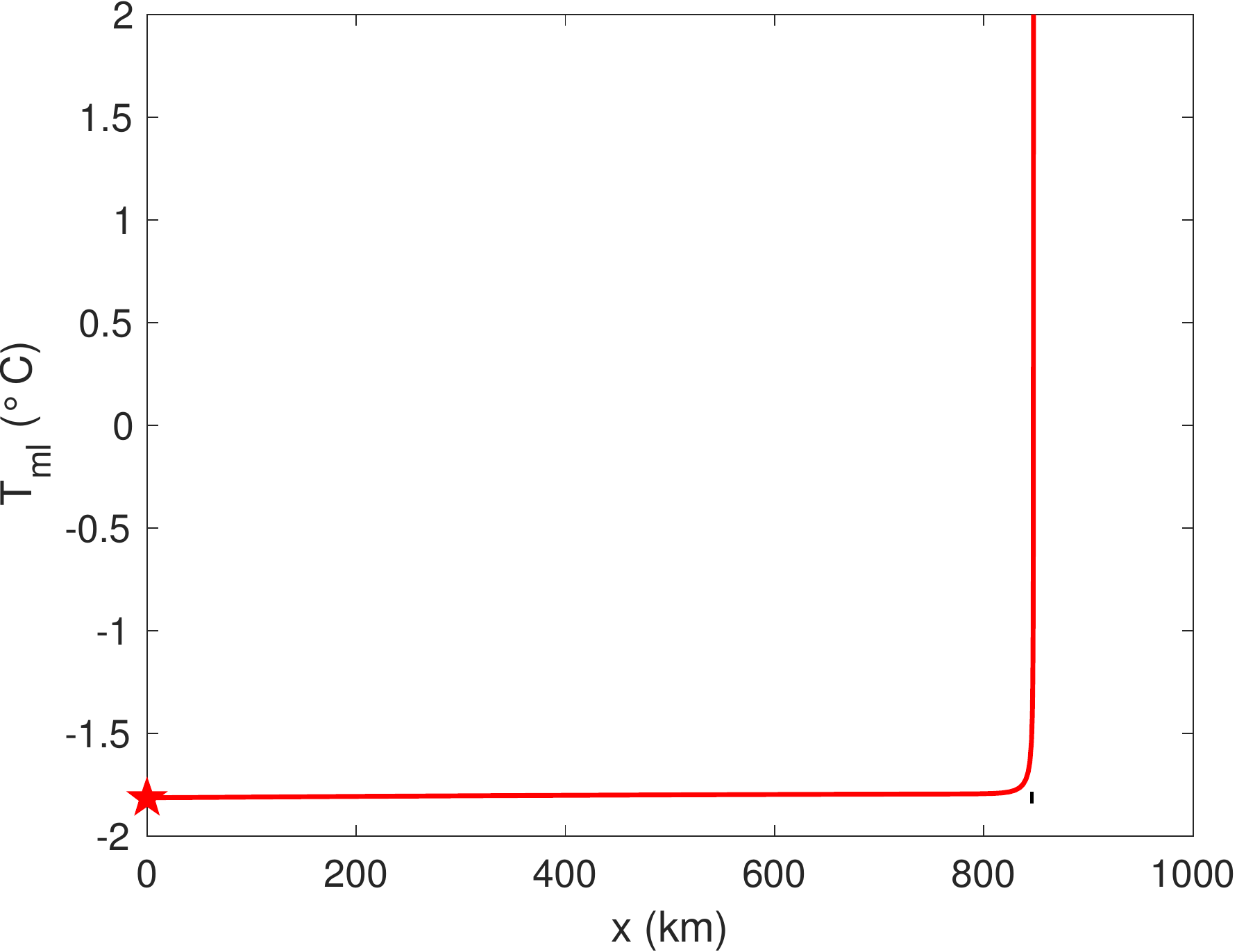}	  	
        (d)\includegraphics[trim = 0 0 0 0, clip, width = 0.21\textwidth]{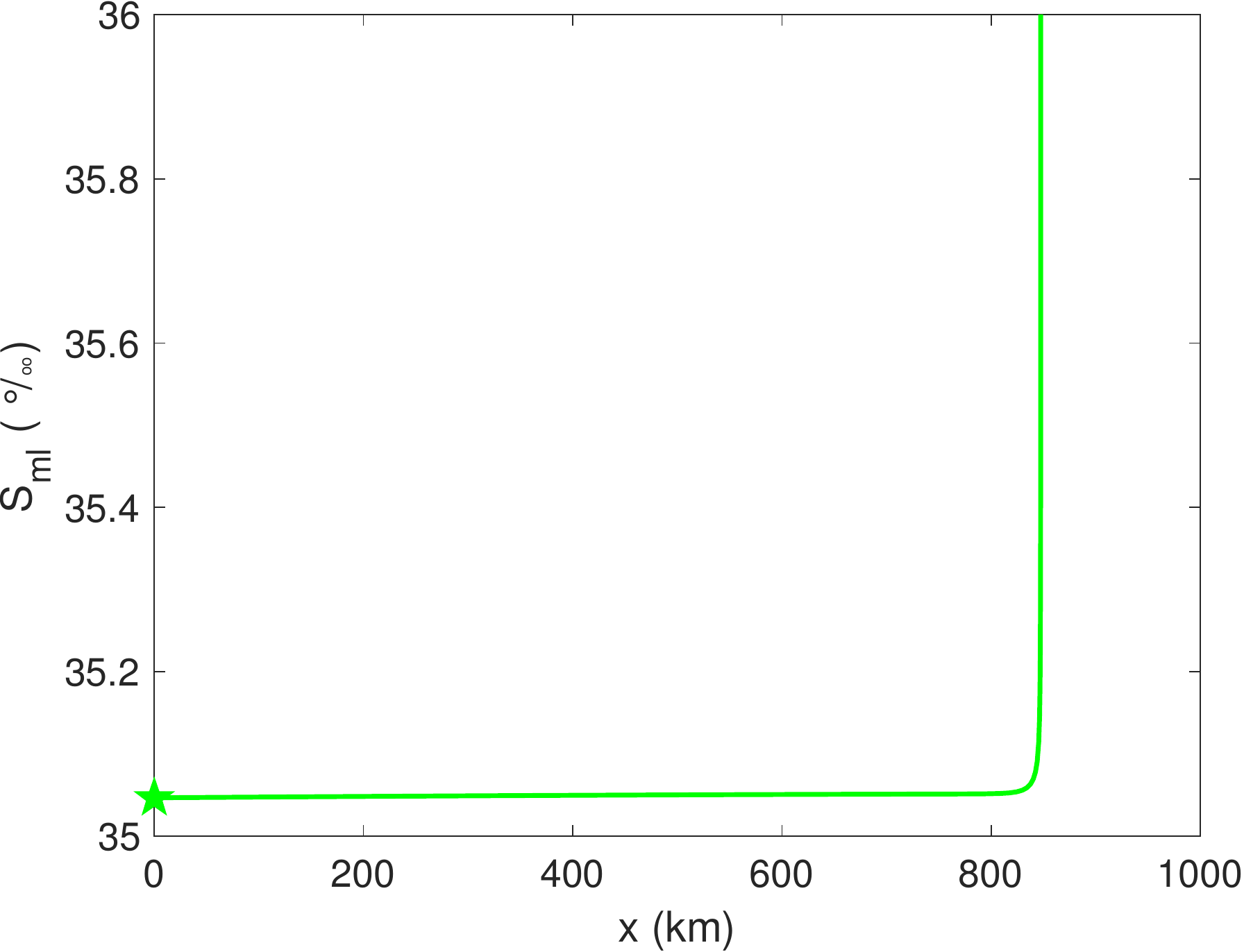}
        
        (e)\includegraphics[trim = 0 0 0 0, clip, width = 0.21\textwidth]{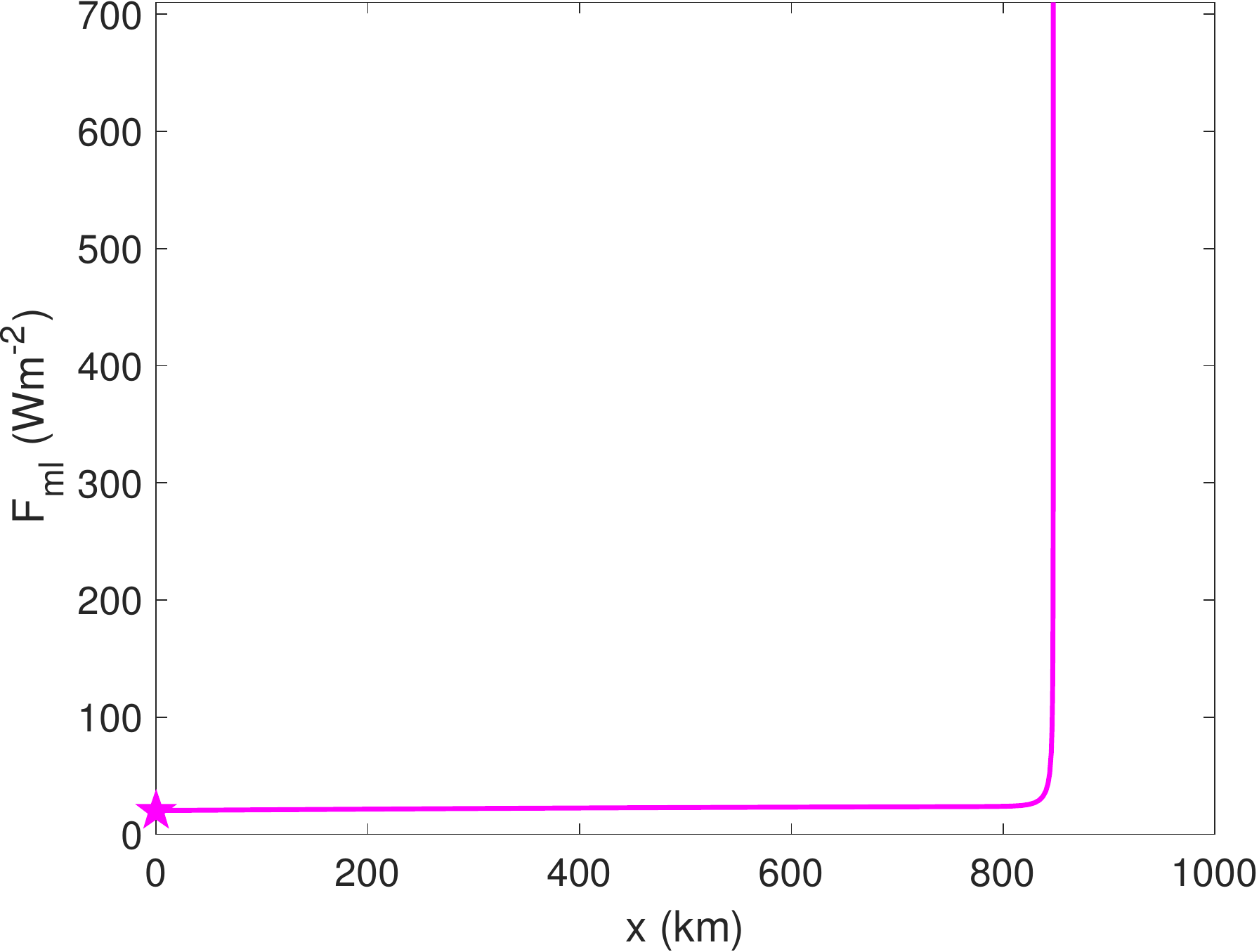}
        (f)\includegraphics[trim = 0 0 0 0, clip, width = 0.21\textwidth]{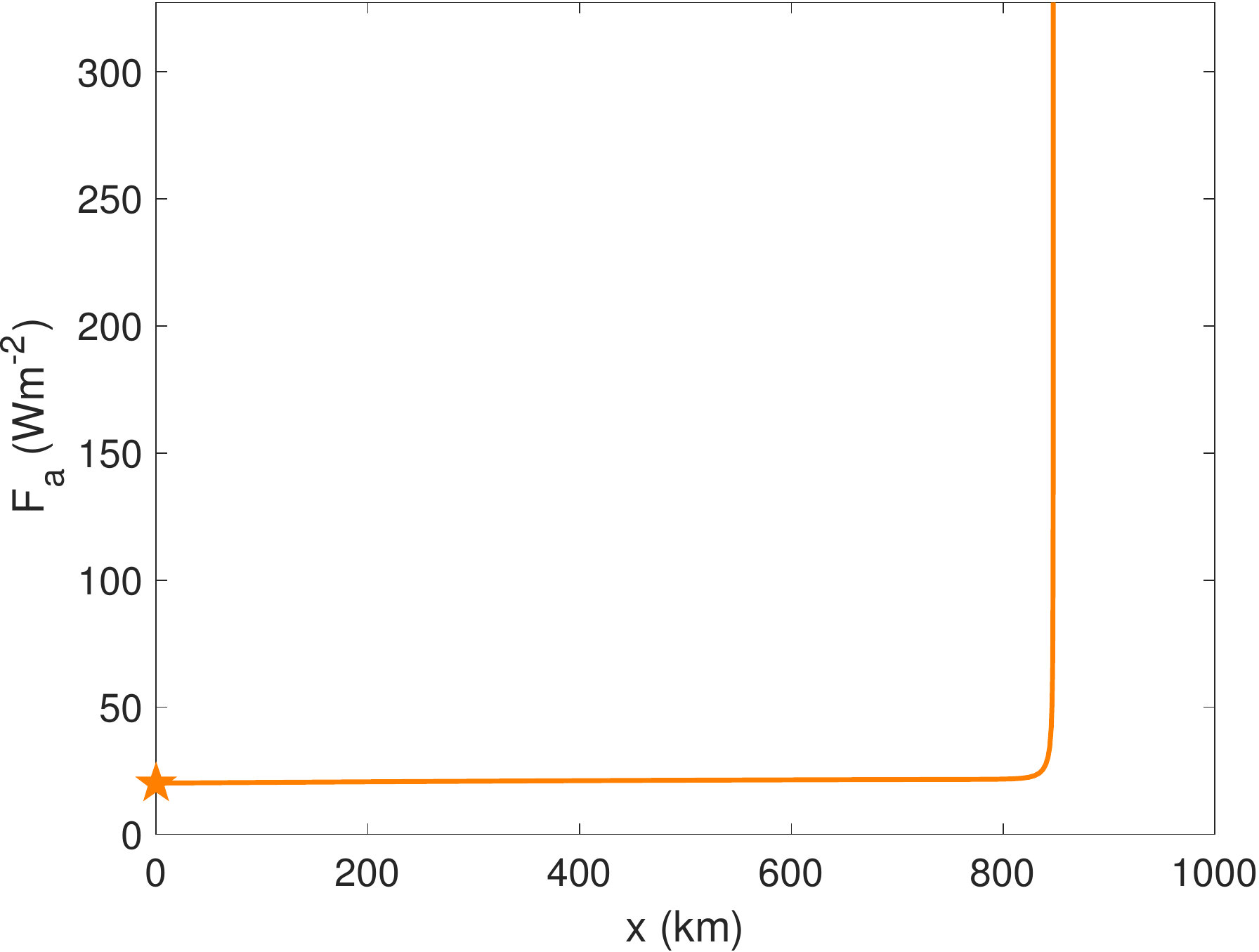}
    \caption{With ice velocity $10^{-3}m/s$ and ocean velocity $2.5\times10^{-1}m/s$. The pentagram on the subplots shows the analytical result. (a)Ice thickness, $h$;(b)mixed-layer (ML) depth, $H$;(c)ML Temp., $T_{ml}$, Thick black line is the \emph{Untersteiner's Wedge}, $\eta = 3.6km$;(d)ML Salinity., $S_{ml}$; (e)mixed-layer heat flux $F_{ml}$; (f) Atmospheric heat flux $F_a$}
    \label{fig:TmlH_h1}
\end{figure}

\begin{figure}[htbp!]
        (a)\includegraphics[trim = 0 0 0 0, clip, width = 0.21\textwidth]{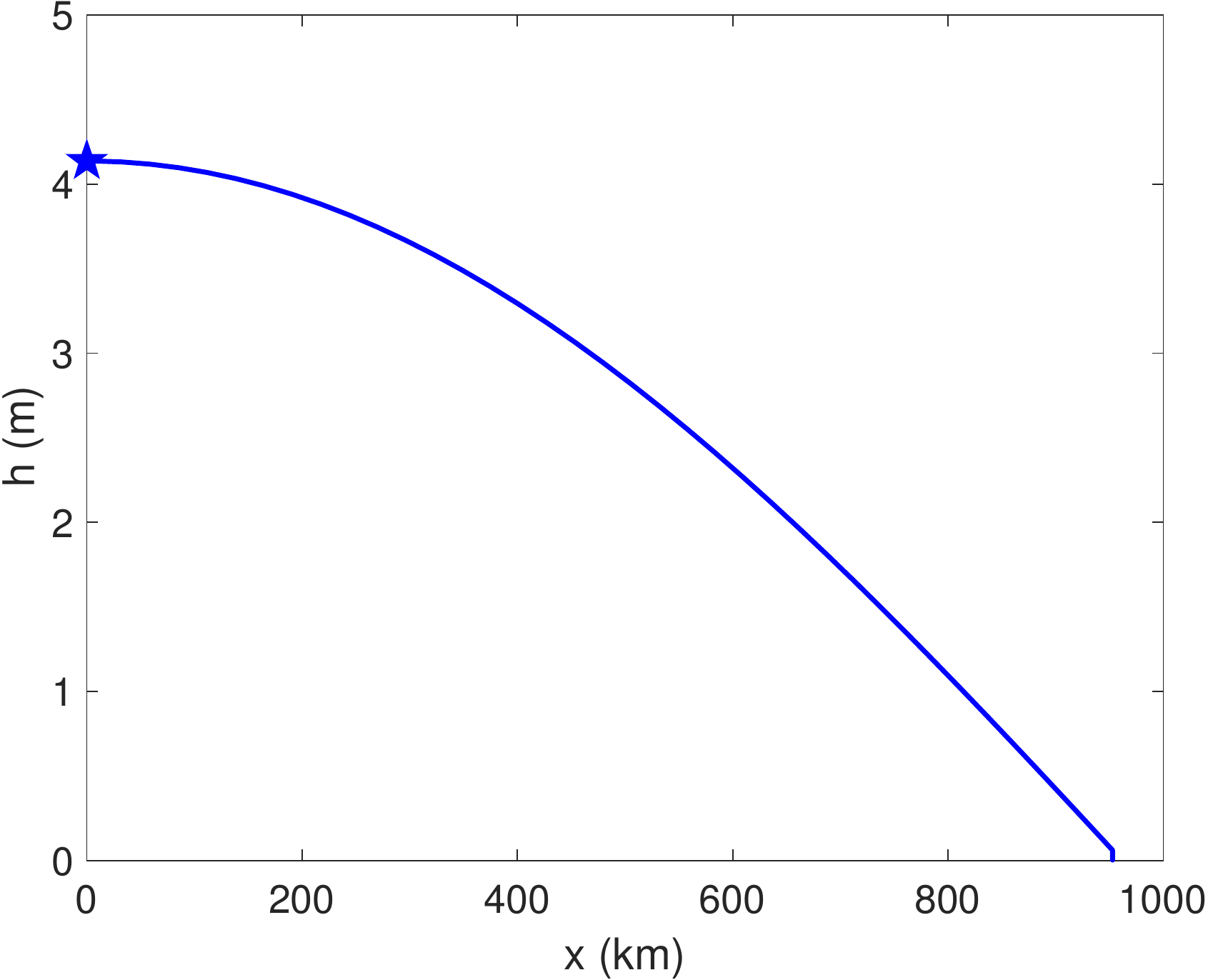}	  	
        (b)\includegraphics[trim = 0 0 0 0, clip, width = 0.21\textwidth]{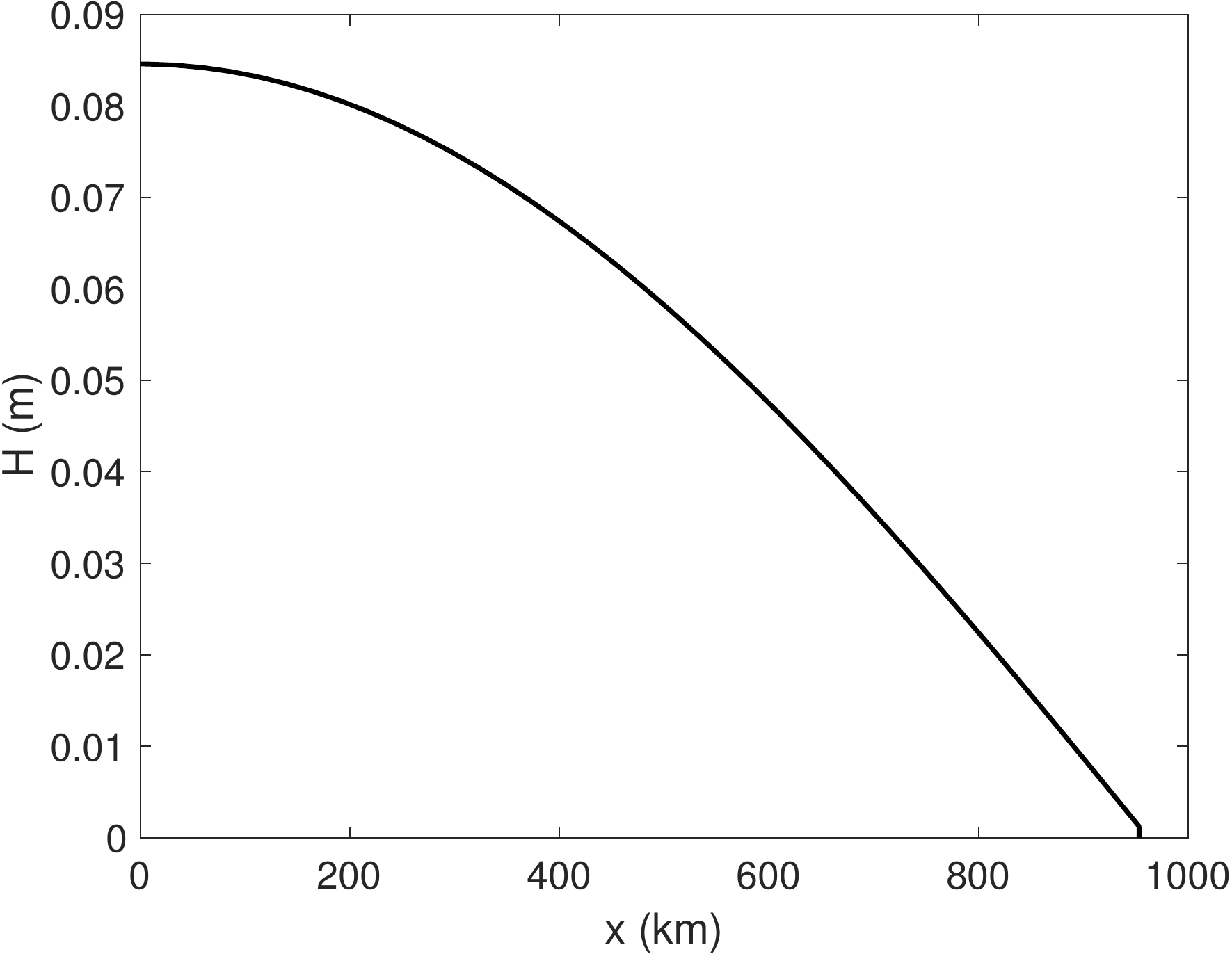}
        	
        (c)\includegraphics[trim = 0 0 0 0, clip, width = 0.21\textwidth]{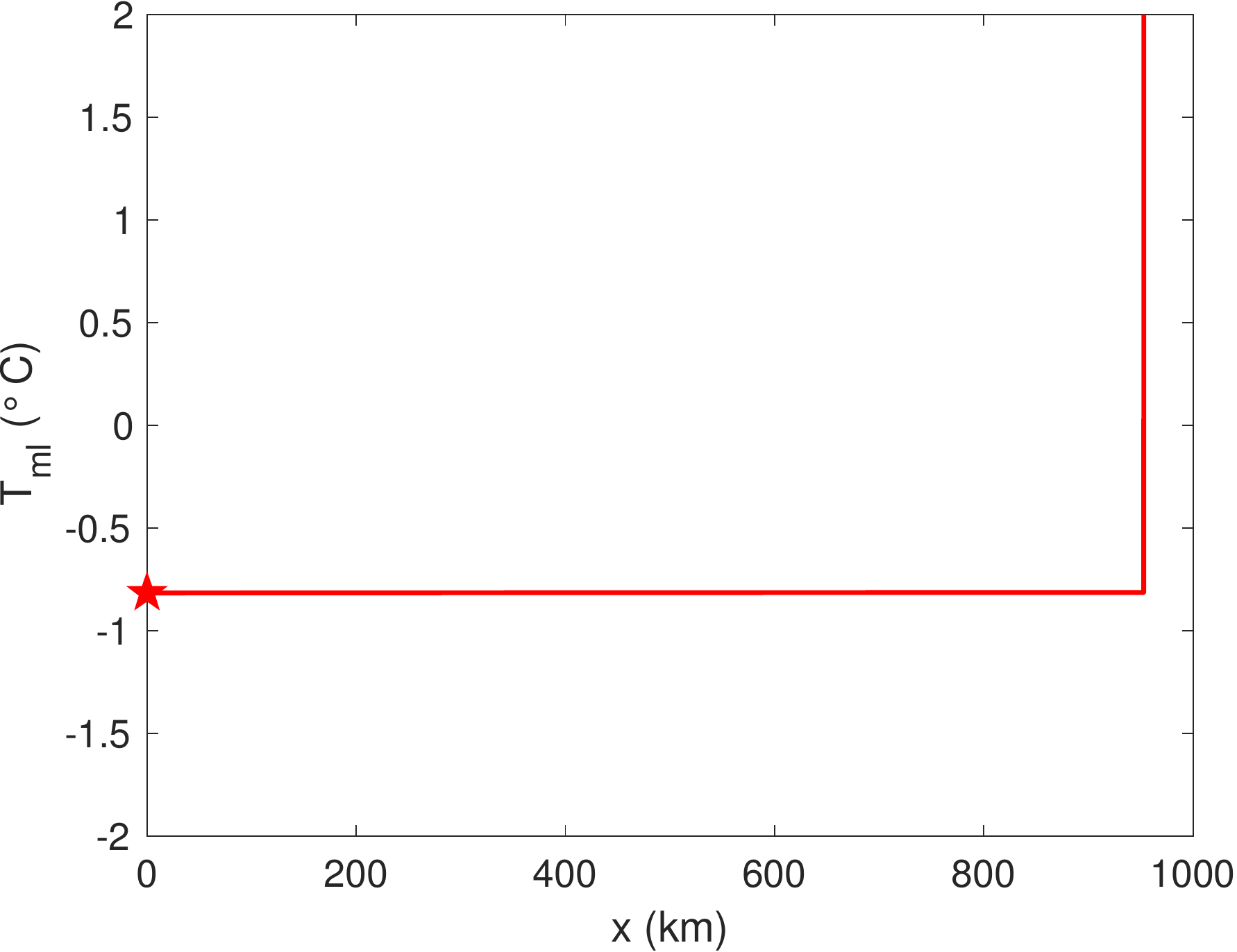}	  	
        (d)\includegraphics[trim = 0 0 0 0, clip, width = 0.21\textwidth]{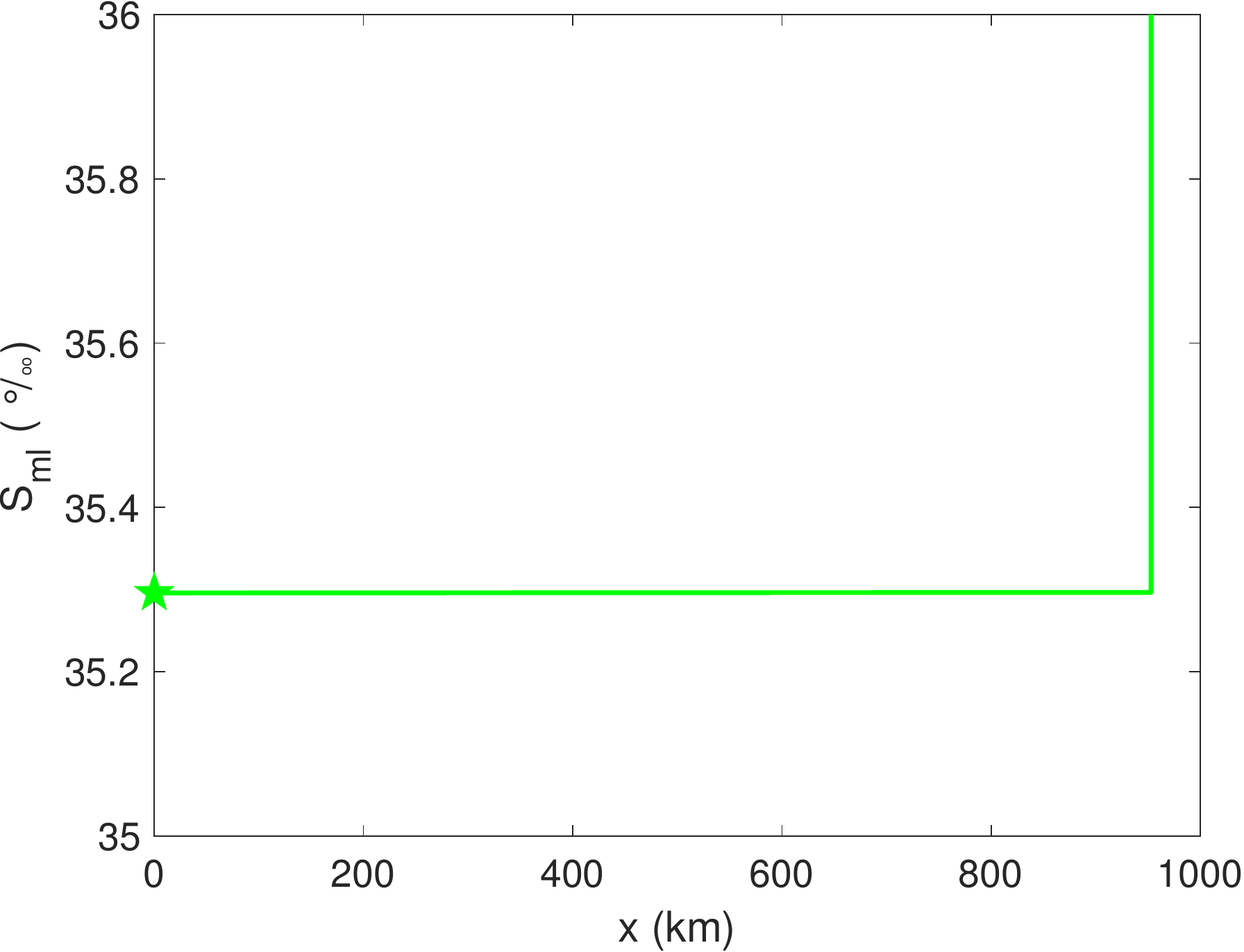}
        
        (e)\includegraphics[trim = 0 0 0 0, clip, width = 0.21\textwidth]{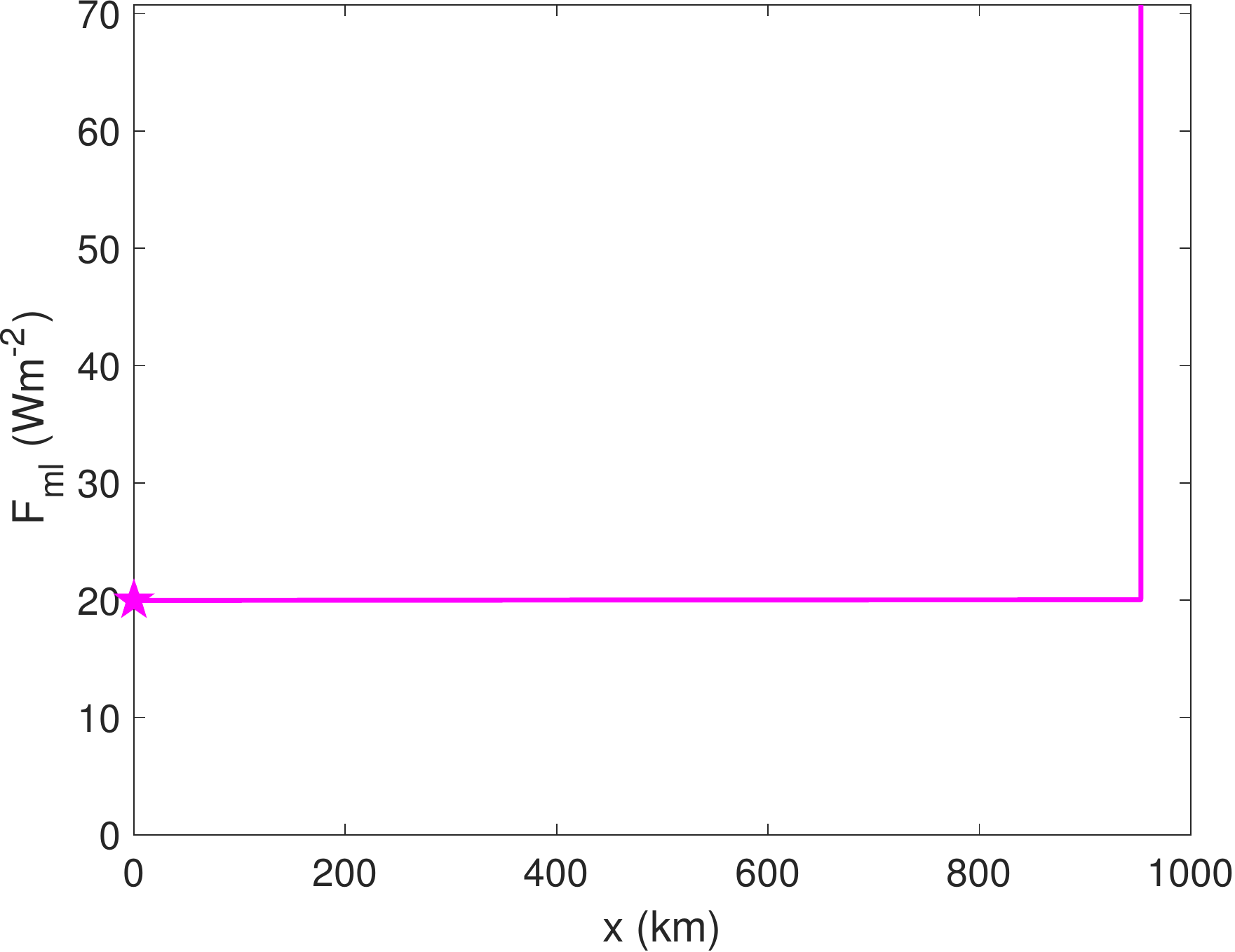}
        (f)\includegraphics[trim = 0 0 0 0, clip, width = 0.21\textwidth]{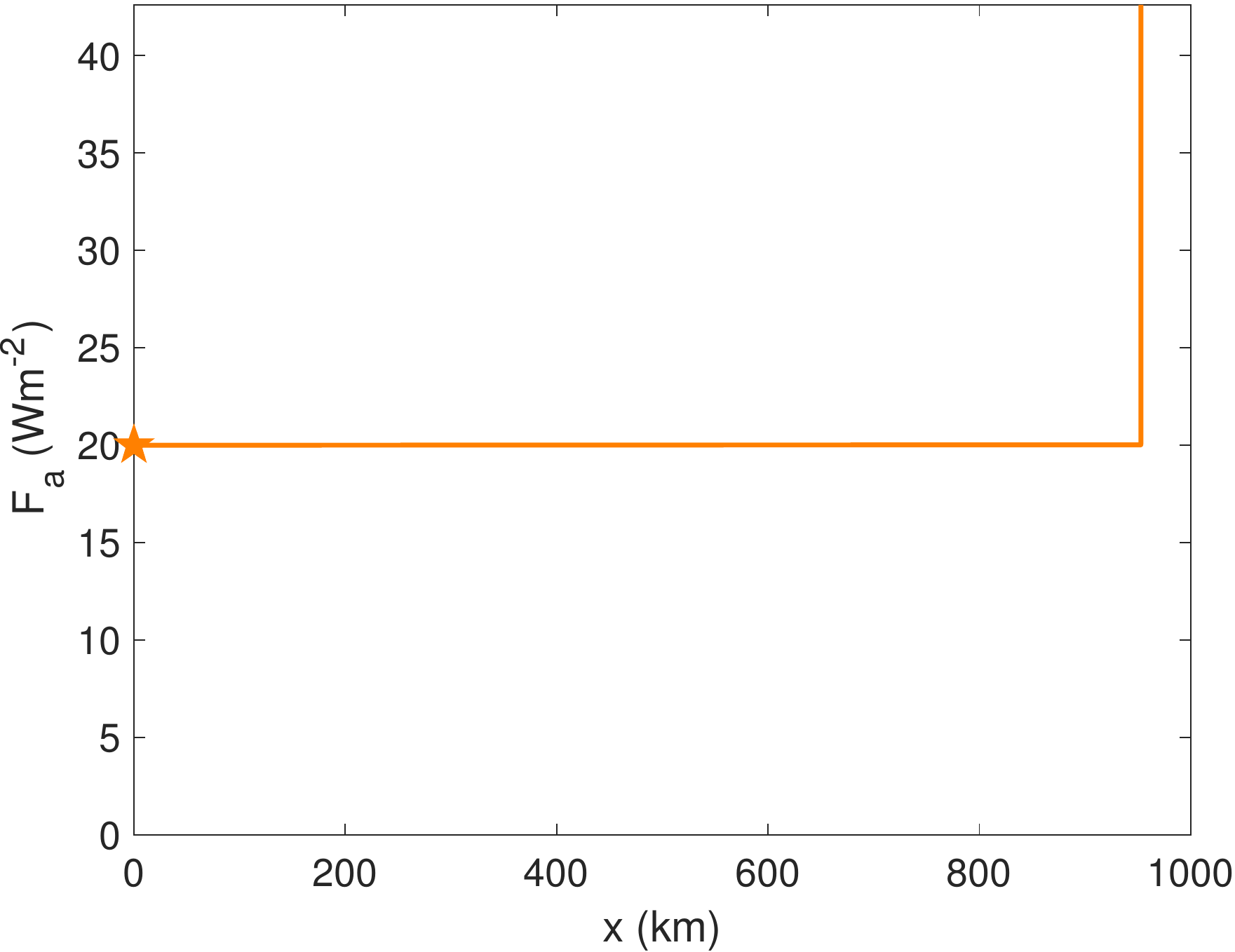}
    \caption{With ice velocity $10^{-5}m/s$ and ocean velocity $2.5\times10^{-2}m/s$. The pentagram on the subplots shows the analytical result. (a)Ice thickness, $h$;(b)mixed-layer (ML) depth, $H$;(c)ML Temp., $T_{ml}$, Thick black line is the \emph{Untersteiner's Wedge}, $\eta = 0.5km$;(d)ML Salinity., $S_{ml}$; (e)mixed-layer heat flux $F_{ml}$; (f) Atmospheric heat flux $F_a$}
    \label{fig:TmlH_h3}
\end{figure}

\begin{figure}[h!]
    \centering
        \includegraphics[trim = 0 0 0 0, clip, width = 0.5\textwidth]{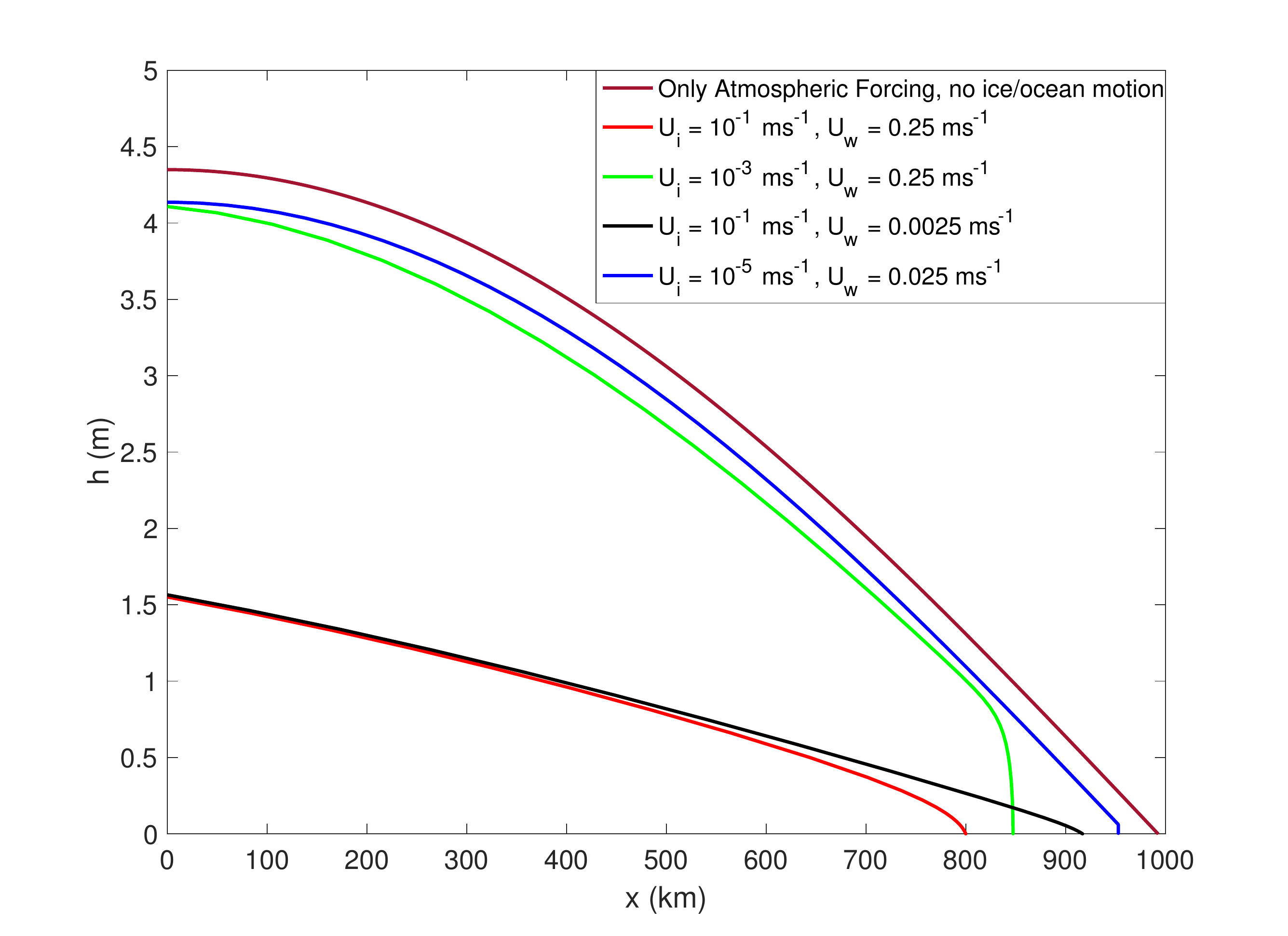}	  	
    \caption{Comparison of sea-ice thickness profiles for the four velocity pairs with the situation if there were only atmospheric forcing. This plot shows how the presence of heat flux from the mixed layer can influence the ice thickness as well as the formation of \emph{Untersteiner's Wedge.}}
    \label{fig:hh_compare}
\end{figure}

\begin{figure}[h!]
    \centering
        \includegraphics[trim = 0 0 0 0, clip, width = 0.5\textwidth]{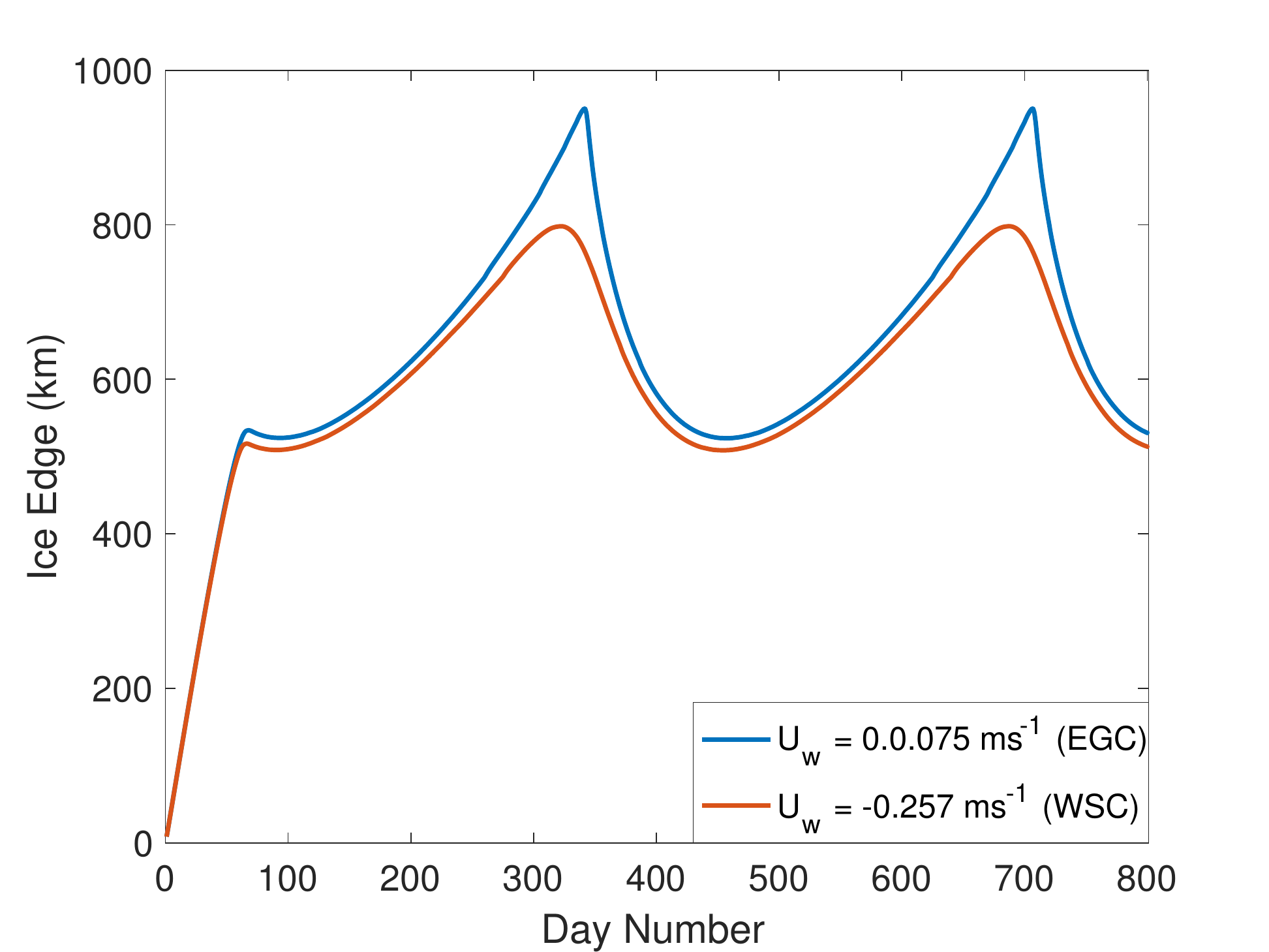}	  	
    \caption{These plots show the time series for daily sea-ice edge for two velocity profiles in EGC and WSC, along with their transient behavior. Two points to note are: (a) both show asymmetrical growth and melt rates, and (b) the time when the two profiles reach their maximum extent are different, and therefore even when the sea ice is retreating northwards in the WSC, it is moving southwards in the EGC.}
    \label{fig:XeTS}
\end{figure}

\begin{figure}[h!]
    \centering
        \includegraphics[trim = 0 0 0 0, clip, width = 0.5\textwidth]{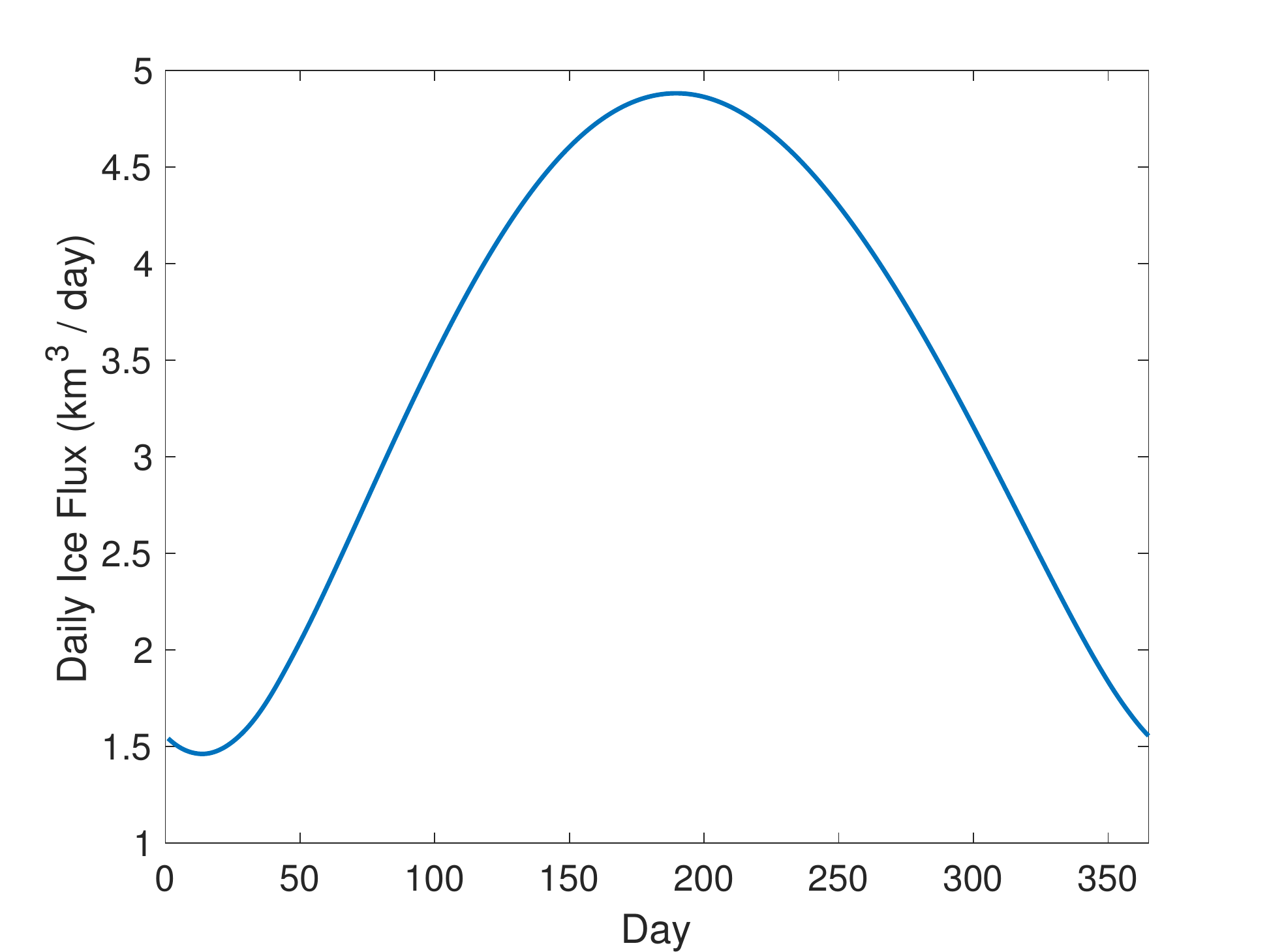}	  	
    \caption{The daily sea-ice flux from the Arctic across the Fram Strait, which converts to $1234km^3/year$.}
    \label{fig:DIF}
\end{figure}
\eject
\subsection*{Movies}

\begin{figure}[h!]
    \centering
    	(a)\includegraphics[trim = 0 0 0 0, clip, width = 0.25\textwidth]{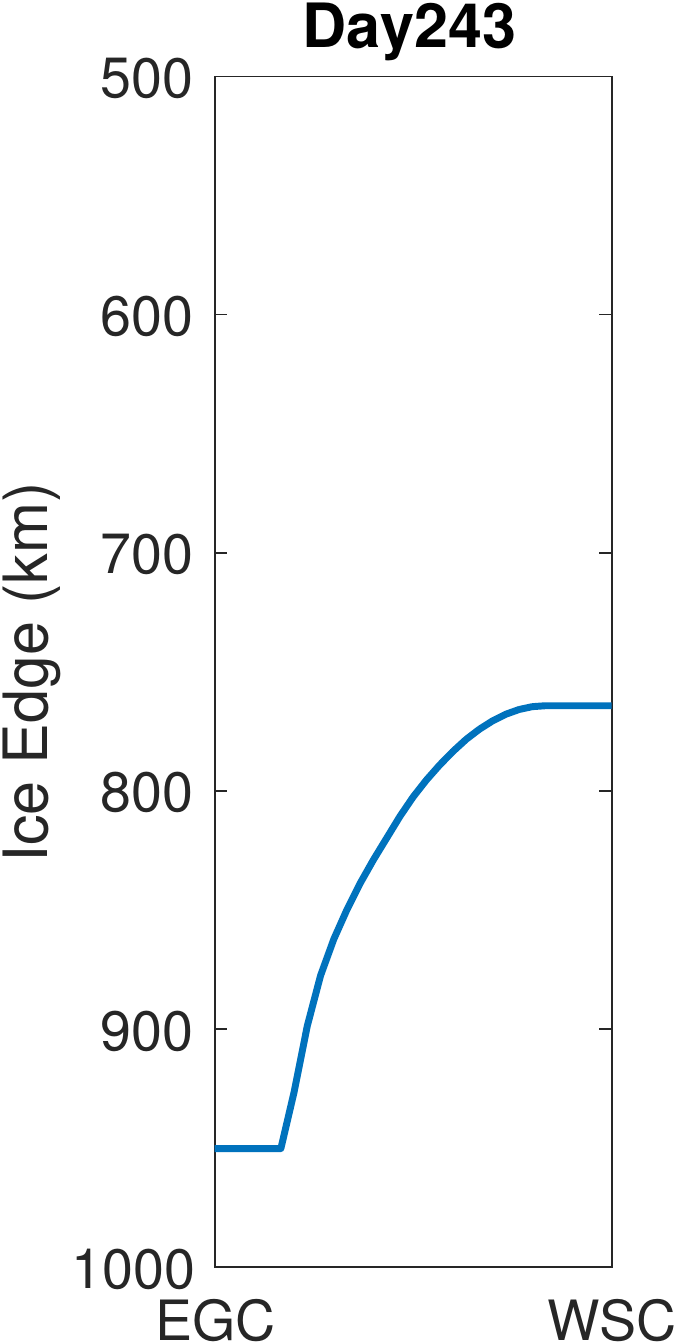}
	(b)\includegraphics[trim = 0 0 0 0, clip, width = 0.25\textwidth]{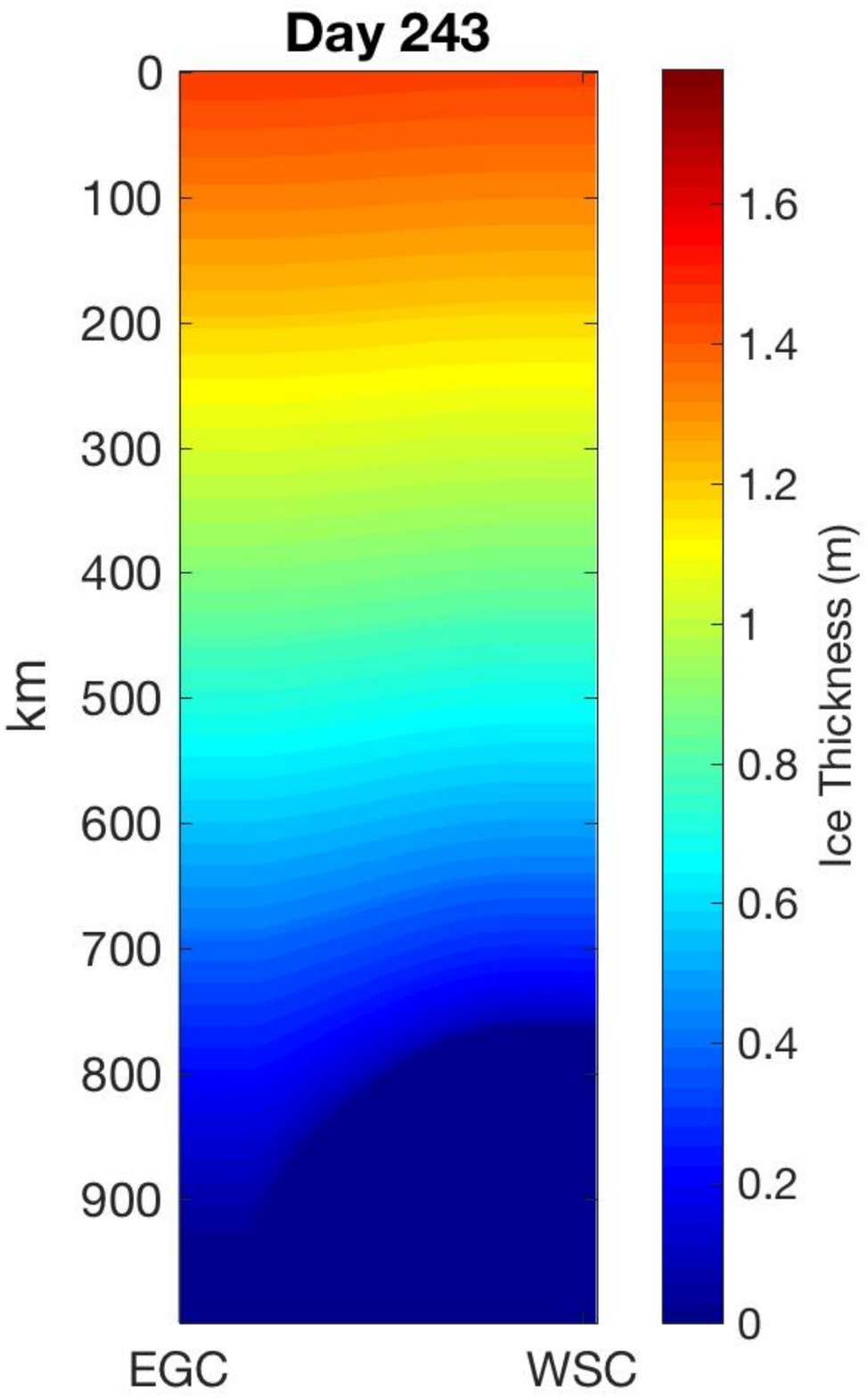}	   	
    \caption{(a) Still image for Movie S1 - This simulation shows the seasonally varying daily sea-ice edge after the transient has been removed for two seasonal cycles. Not only is there an asymmetry between the growth and melt rates of sea-ice edge but also an asymmetry between when different velocity profiles reach their maximum extent. (b) Still image for Movie S2 - This simulation shows the daily flow of sea ice across the Fram Strait, with the thickness variation shown in colors. The sea-ice edge here corresponds to that in Movie S1.}
    \label{fig:Movie1}
\end{figure}
\eject

\end{document}